\documentclass[epj-spec]{svjour}
\usepackage{graphics}
\usepackage{graphicx}
\begin{document}
\title{Gluon polarisation from high transverse momentum hadron pairs production @ COMPASS}
\subtitle{}
\author{Lu\'is Silva \thanks{{\it For correspondence:
    }~\email{lsilva@lip.pt}} \\ LIP}
\institute{On behalf of the COMPASS collaboration}
\abstract{
A new preliminary result of a gluon polarisation $\Delta G/G$ obtained selecting high transverse momentum
hadron pairs in DIS events with $Q^2>1 \ (\mbox{GeV/}c)^2$ is
presented. Data has been collected by COMPASS at CERN during the 2002-2004 years. In the extraction of $\Delta G/G$ contributions coming from
the leading order $\gamma q$ and QCD processes are taken into
account. A new weighting method based on a neural network approach is
used. Also a preliminary result of $\Delta G/G$ for events with $Q^2<1
\ (\mbox{GeV/}c)^2$ is presented.
} 
\maketitle
\section{Introduction}
\label{intro}

Deep inelastic scattering (DIS) of leptons on
nucleons is an important tool to reveal the inner
structure of the nucleon. The DIS experiment at SLAC in the 60's
showed the scaling predicted by J. Bjorken, in the limit $Q^2
\rightarrow \infty$; this discovery was
celebrated in 1990 by the Nobel Prize in Physics awaded to
J.I. Friedman, H.W. Kendal and R.E. Taylor. The observation of this
scaling was the first evidence of point-like constituents inside
the nucleon.

DIS of \emph{polarised} leptons on \emph{polarised} nucleons is a tool
to study the \emph{spin} structure of the
nucleon. The first experiments using polarised ep scattering were
perfomed by the E80 \cite{Alguard:1976bm} and E130 \cite{Baum:1983ha} Collaborations at SLAC, measuring the
spin-dependent asymmetries with a significant value consistent with the Ellis-Jaffe sum
rule \cite{ellis-jaffe}. Surprisingly on 1987 the EMC experiment at CERN, with an
extended kinematic range down to $0.01 < x$, announced that,
contradicting previous results and predictions, the measured
quark contribution to nucleon spin is small ($0.12 \pm 0.17$)
\cite{emc} and its result has been confirmed by other experiments
\cite{Abe:1997cx,Abe:1998wq,Anthony:1996mw,Anthony:1999py,Adams:1997tq,Airapetian:1998wi}. In 2007, COMPASS
collaboration measured this contribution with a higher precision \cite{comp.del_sigma},
using a NLO QCD fit with all world data available including 43 points
measured by COMPASS, confirms that approximately 1/3 of the spin is
carried by the quarks.

Since the quark contribution does not account fully for the nucleon spin
some contributions need to be found to solve this ``spin
crisis''. As nucleons are also made of gluons together with quarks,
the most natural would be to include the contributions from the gluons
and from orbital angular momentum.

Thus the nucleon spin can be written as:
\begin{equation}
\frac{1}{2}=\frac{1}{2}\Delta \Sigma + \Delta G + L
\end{equation}

$\Delta \Sigma$ and $\Delta G$ are, respectively, the quark helicity and gluon contributions to the nucleon spin and $L$ is the contribution to the nucleon spin coming from orbital angular momentum from the partons (quarks and gluons).

The aim of this study is to estimate the gluon polarisation using the high
transverse momentum (high $p_T$) hadron pairs. The analysis is
performed in two complementary kinematic regions: $Q^2 <1 \ (\mbox{GeV}/c)^2$
(low $Q^2$ region) and $Q^2 >1 \ (\mbox{GeV}/c)^2$ (high $Q^2$ region). The
present work is mainly focused on the analysis for high $Q^2$. However,
the analysis at low $Q^2$ region is summarised in section \ref{sec:lowq2}.

\section{Analysis Formalism}
\label{sec:analysis}

Spin-dependent effects can be measured experimentally using the helicity asymmetry 

\begin{equation}
  A_{\rm LL} = \frac{\Delta \sigma}{2 \sigma}=
  \frac{\sigma^{\uparrow \Downarrow} - \sigma^{\uparrow \Uparrow}
  }{\sigma^{\uparrow \Downarrow} + \sigma^{\uparrow \Uparrow} } \label{eq:asy}
\end{equation}
defined as the ratio of polarised ($\Delta \sigma$) and unpolarised ($\sigma$)
cross sections. $\uparrow \Uparrow$ and $\uparrow \Downarrow$ refer to
the parallel and anti-parallel configuration of the beam lepton
spin ($\uparrow$) with respect to the target nucleon spin ($\Uparrow$ or $\Downarrow$).

According to the factorisation theorem, the (polarised) cross
sections can be written as 

\begin{eqnarray}
\sigma = \sum_i e^2_i q_i \otimes \hat{\sigma} \otimes D \label{sigma}\\
\Delta \sigma = \sum_i e^2_i \Delta q_i \otimes \Delta \hat{\sigma} \otimes D \label{eq:del_sigma}
\end{eqnarray}
i.e. the convolution of the parton distribution
functions, ($\Delta$)$q_i$, the hard scattering partonic cross section,
($\Delta$)$\hat{\sigma}$, and the fragmentation function $D$.

\begin{figure}
  \begin{center} 
    \resizebox{0.8\columnwidth}{!}{%
      \includegraphics{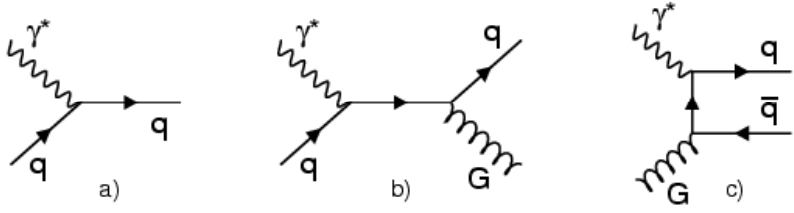}}
  \end{center}
\caption{The contributing processes: a) DIS LO, b) QCD Compton and c) Photon-Gluon Fusion.}
\label{fig:procs}       
\end{figure}

The gluon polarisation is measured directly via the Photon-Gluon
Fusion process (PGF); which allows to probe the gluon inside the nucleon. Two other processes
compete with the PGF process in the leading order QCD approximation, namely
the virtual photo-absorption leading order (LO) process and the gluon
radiation (QCD Compton) process. In Fig. \ref{fig:procs} all contributing processes are depicted.

The helicity asymmetry for the high $p_T$ hadron pairs in high $Q^2$
regime can be written as:

\begin{equation}
A_{\rm LL}^{2h}(x_{Bj})= R_{\rm PGF} \, a_{\rm LL}^{\rm PGF}\frac{\Delta G}{G}(x_G) +
R_{\rm LO} \, D \, A_1^{\rm LO}(x_{Bj}) + R_{\rm QCDC} \, a_{\rm LL}^{\rm QCDC}
A_1^{\rm LO}(x_C) \label{eq:allmain}
\end{equation}

The $R_i$ (the index $i$ refers to the different processes) are the fractions of each process. $a_{\rm LL}^i$ represents the
partonic cross section asymmetries, $\Delta\hat{\sigma}^i/\hat{\sigma}^i$, (also known as analysing power). $D$
is the depolarisation factor\footnote{The Depolarisation factor is the fraction of the muon beam polarisation transfered to the virtual photon.}. The virtual photon asymmetry $A_1^{\rm LO}$ is defined as 
\begin{equation}
A_1^{\rm LO} \equiv \frac{\sum_i e_i^2 \Delta q_i}{\sum_i e_i^2 q_i}.
\end{equation}

To extract $\Delta G/G$ from eq. (\ref{eq:allmain}) the contribution
from the physical background processes LO and QCD Compton need to be
estimated. This is done using MC simulation to calculate $R_i$
fractions and $a_{\rm LL}^i$. The virtual photon asymmetry $A_1^{\rm LO}$ was
estimated using a parametrisation based on inclusive the $A_1$ asymmetry
data \cite{compassrho}.

For the inclusive asymmetry $A_{\rm LL}^{incl}$ a similar decomposition as
eq. (\ref{eq:allmain}) can be applied:

\begin{equation}
A_{\rm LL}^{incl}(x_{Bj})= R_{\rm PGF}^{incl}
a_{\rm LL}^{incl,{\rm PGF}}\frac{\Delta G}{G}(x_G) + R_{\rm LO}^{incl} D
A_1^{\rm LO}(x_{Bj}) + R_{\rm QCDC}^{incl} a_{\rm LL}^{incl,{\rm QCDC}}
A_1^{\rm LO}(x_C).
 \label{eq:incl}
\end{equation}

Note that $y$, $D$\footnote{Depolarisation factor is the fraction of polarisation
  transferred from the in coming muon to the virtual photon}, $x_{Bj}$, $x_G$ and $x_C$ in the inclusive
and high $p_T$ sample can be different. The extraction  of $\Delta G/G$ requires a new
definition of  the averaged $x_G$ at which the measurement is
performed:
\begin{equation}
x_G^{av}=\frac{\alpha_1 x_G - \alpha_2 x_G'}{\beta}
\end{equation}
where:
\begin{eqnarray}
\alpha_1 &=& a_{\rm LL}^{\rm PGF} R_{\rm PGF} - a_{\rm LL}^{incl,{\rm PGF}} R_{\rm LO}\frac{R_{\rm PGF}^{incl}}{R_{\rm LO}^{incl}} \\
\alpha_2 &=& a_{\rm LL}^{incl,{\rm PGF}}
R_{\rm QCDC}\frac{R_{\rm PGF}^{incl}}{R_{\rm LO}^{incl}}
\frac{a_{\rm LL}^{\rm QCDC}}{D}\\
\beta &=& \alpha_1-\alpha_2.
 \label{eq:form:alfa}
\end{eqnarray}

The definition of $x_G^{av}$ relies on the assumption of linear
dependence of $\Delta G/G$ on $x_G$. This assumption is well justified
by the narrow $x_G$ bin used.

Using eq. (\ref{eq:allmain} and \ref{eq:incl}) and neglecting small
terms the following
expression is obtained:

\begin{eqnarray}
\frac{\Delta G}{G}(x_G^{av}) &=& \frac{A_{\rm LL}^{2h}(x_{Bj})+A^{corr} }{\beta}\nonumber\\
A^{corr}&=& - A_1(x_{Bj})D \frac{R_{\rm LO}}{R_{\rm LO}^{incl}} - A_1(x_C)
\beta_1 + A_1(x_C')\beta_2
 \label{eq:form:gluon}
\end{eqnarray}
and
\begin{eqnarray}
\beta_1 &=& \frac{1}{R_{\rm LO}^{incl}}\bigg[a_{\rm LL}^{\rm QCDC} R_{\rm QCDC} -
a_{\rm LL}^{incl,{\rm QCDC}} R_{\rm QCDC}^{incl}\frac{R_{\rm LO}}{R_{\rm LO}^{incl}}\bigg]\nonumber\\
\beta_2 &=& a_{\rm LL}^{incl,{\rm QCDC}}
\frac{R_{\rm QCDC}^{incl}}{R_{\rm LO}^{incl}}\frac{R_{\rm QCDC}}{R_{\rm LO}^{incl}}\frac{a_{\rm LL}^{\rm QCDC}}{D}
\label{eq:form:betas}
\end{eqnarray}

The term $A^{corr}$ comprises the correction due to the other two
processes; namely the LO and the QCD Compton processes. $\beta_1$, $\beta_1$,
$\beta_1$, $x_C$ and $x_G^{av}$ are estimated using high $p_T$ and
inclusive MC samples. The gluon polarisation $\Delta G/G$ is extracted using a weight evaluated in a event-by-event analysis describe in section \ref{deltag}.

\section{COMPASS Experiment}
\label{sec:compass}
COMPASS is a deep inelastic scattering experiment located at the Super
Proton Synchrotron (SPS) accelerator at CERN. It is dedicated to the study
of the spin structure of the nucleon and to hadron spectroscopy. The
experimental setup consists in three main components: a polarised muon beam, a
polarised target and a two-stage spectrometer.

A $400 \ \mbox{GeV/}c$ proton beam extracted from the SPS collides on a
beryllium target producing mainly $\pi$ and $K$ mesons. Which are transported through a 600 m long
decay channel. Due to the parity violation in weak decays of the
parent hadrons ($\pi,\ K \rightarrow \mu \ \nu_\mu$) the newly
produced muons are naturally polarised at the energy of 160 GeV.

\begin{figure}
  \begin{center} 
    \resizebox{0.5\columnwidth}{!}{%
      \includegraphics{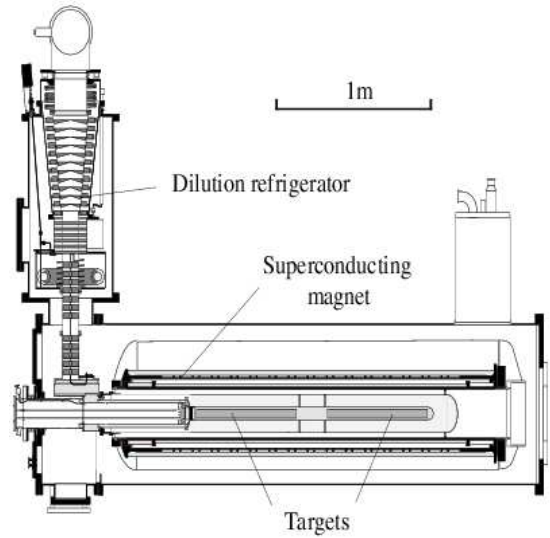}}
  \end{center}
\caption{Schematic drawing of the polarised target.}
\label{fig:targ}       
\end{figure}

The polarised target is composed by two cylindrical target cells
polarised in opposite directions, filled with deuterated lithium
($^6$LiD) solid state material. Each cell is 60 cm long and has a
radius of 3 cm. They are disposed longitudinally one after the other
with a separation of 10 cm and embedded in a superconducting solenoid
magnet that provides a very homogeneous field of 2.5 Tesla. The target
cells are kept under a temperature below 60 mK. This solenoid magnet was used in the SMC
experiment and has a geometric acceptance of $\approx  \pm 70$ mrad. In
Fig. \ref{fig:targ} a drawing of the polarised target is shown. 

The polarisation method is based on the dynamic nuclear polarisation
(DNP) technique \cite{Abragam:89a}; the paramagnetic centers, the electrons, of the
target cells under the high and homogeneous solenoid magnetic field and at a very low temperature are polarised to high
degree. A microwave field is applied to the target material to
transfer the polarisation from the paramagnetic centers to the
nucleons. Since the desired situation is to have two cells oppositely
polarised two independently microwave systems are required. Thus
depending on the microwave frequency the nucleon spins of the cells
can be polarised parallel or anti-parallel with respect to the beam polarisation. A dipole magnet with a field of 0.5 T
perpendicular to the solenoid magnetic field performs the
spins rotation of the target cells with respect to beam
direction. This field rotation of $180^{\rm o}$ is done in a regular
basis to minimise systematics errors due to geometrical acceptance of the
solenoid.

\begin{figure}[tbp] 
\centerline{\includegraphics[clip,width=0.9\textwidth]{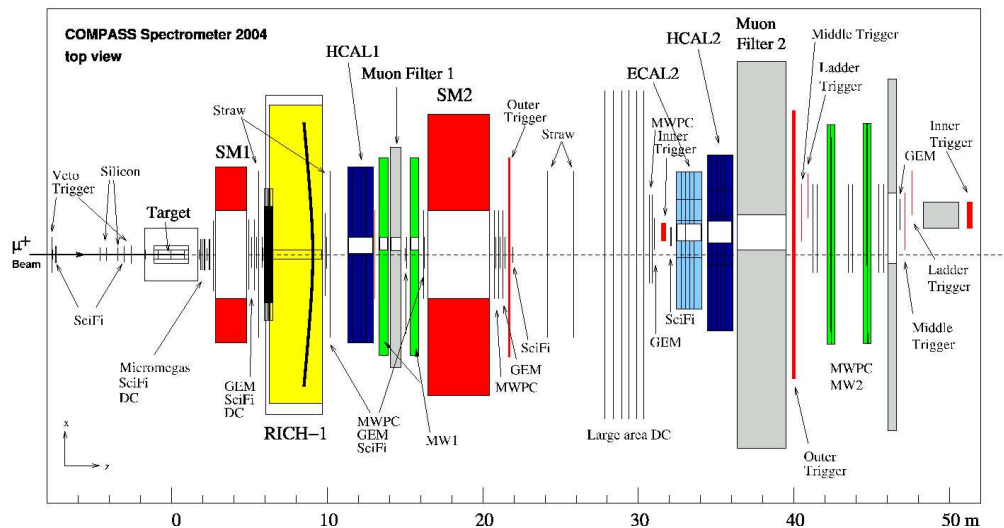}} 
  \caption{Compass 2004 muon setup top 
    view.}  
  \label{fig:layout.setup} 
\end{figure}

The COMPASS spectrometer covers a large kinematic region ($10^{-4}
~(\mbox{GeV}/c)^2<Q^2<60 \ (\mbox{GeV}/c)^2$, $10^{-5}<x_{Bj}<0.5$). Each stage spectrometer is composed by a magnet,
tracking chambers and trigger devices. The first spectrometer is
disposed downstream after the polarised target, it covers an
acceptance of $\pm 180$ mrad and has a bending magnet power of 1
Tm. Therefore this spectrometer is mainly devoted to low momentum
particles, it is also known as large angle spectrometer (LAS). The
next spectrometer is the small angle spectrometer (SAS) and devoted to
high momentum particles, it covers an acceptance of $\pm 30$ mrad,
with a bending power of 4.4 Tm. The tracking system is distributed in
both stage spectrometers and it can be divided in three main
zones: very small area trackers (VSAT) --the set of tracking planes
between the solenoid magnet and the LAS magnet-- , the
small area trackers (SAT)  --the set of tracking planes between the LAS
and SAS magnets-- and large area trackers (LAT) --the set of planes
after the SAS magnet--. In Fig. \ref{fig:layout.setup} all regions and components of the
COMPASS spectrometer are illustrated.

For a more
complete description of the experimental apparatus the reader is
addressed to \cite{compass}.

\section{Data Selection}
\label{sec:data}

The data sample used in this analysis includes data from 2002, 2003
and 2004 years. The selected events have a primary vertex containing
an incoming beam muon, a scattered outgoing muon and at least two
outgoing hadrons with high transverse momentum.

The following kinematic cuts are applied: $Q^2>1 \ (\mbox{GeV}/c)^2$. A cut applied is on the fraction of energy taken by the virtual photon, $y$: $0.1 < y < 0.9$; events with $y <
0.1$ are rejected because their depolarisation
factor is rather low, while
  events with $y > 0.9$ are rejected because they are
  strongly affected by radiative effects, which are difficult to
  evaluate.

The incoming muon, $\mu$, is required to cross both target cells, to
ensure the same flux. Two particles with highest $p_{T}$ associated with the primary vertex
besides the $\mu$ and
$\mu^\prime$ are considered as {\em hadron candidates}. They must
fulfil the following requirements:
\begin{itemize}

\item The hadron candidates are not muons. There is indeed a small
probability
  for a pile-up muon to be included in the primary vertex and therefore being
  considered as a hadron candidate. When the energy measurement by the
  calorimeters is available, the hadron candidate is rejected if
  $E_{cal}/p < 0.3$, where $E_{cal}$ is the total energy measured by
  the hadronic calorimeters associated to the track of momentum $p$. 

\item The quality of the track reconstruction is good. 

\item Hadrons do not go through the solenoid. The hadron tracks
are
  extrapolated to the entrance of the solenoid and then the distance
  between the track and the z axis should be less than the radius of
  the solenoid aperture.
\end{itemize}

The following cuts are applied to the leading (highest transverse
momentum) and sub-leading hadrons:
\begin{itemize}
\item Both hadrons must have a transverse momentum above $0.7 \ \mbox{GeV/c}$. This requirement constitutes the  high $p_T$ cut.
\item $x_F > 0$, $z > 0$ and $z_1 + z_2 < 0.95$. This last cut is
  meant to reject events from exclusive  production.
\item Invariant mass of the two high $p_T$ hadrons must be greater
than
  $1.5 \ (\mbox{GeV/c})^2$. This cut is intended to remove the
  virtual photon events which fluctuate to a vector meson such as a
  $\rho$, that 
  afterwards  decay into two hadrons.
\end{itemize}

The number of events and the percentage  that survives
each cut are displayed in Table~\ref{tab:data:cuts}.
In this table, the event {\em
candidates} are events that pass all kinematic and high $p_T$ cuts.
PID hadrons refer to event which pass the first cut of
{\em hadrons candidates}.

\begin{table}[!hbp]
\begin{center}
\begin{tabular}{|c||c|c|c|c|c|}
\hline
 &\multicolumn{4}{c|}{ number of events that survive each cut}&\\
\cline{2-6}
Cuts & 2002& 2003& 2004& All years& \% \\
\hline\hline
Event candidate                          &89111 &309893 &524862 &923866 &100.0\\
Invariant  mass                      &59711 &208055 &350989 &618755 &67.0\\
PID hadrons                          &52363 &180965 &301698 &535026 &57.9\\
$x_{F}>0$, $z>0$                     &51325 &176426 &294970 &522721 &56.6\\
$x_{Bj}>0$, $z_1+z_2<0.95$           &49962 &172431 &288732 &511125 &55.3\\
Hadron quality                           &49585 &170943 &286685 &507213 &54.9\\
\hline
\end{tabular}
\end{center}
\caption{Table summarising cuts.} \label{tab:data:cuts}
\end{table}

\begin{figure}[tbp]
\begin{center}
  \resizebox{0.32\textwidth}{!}{
    \includegraphics{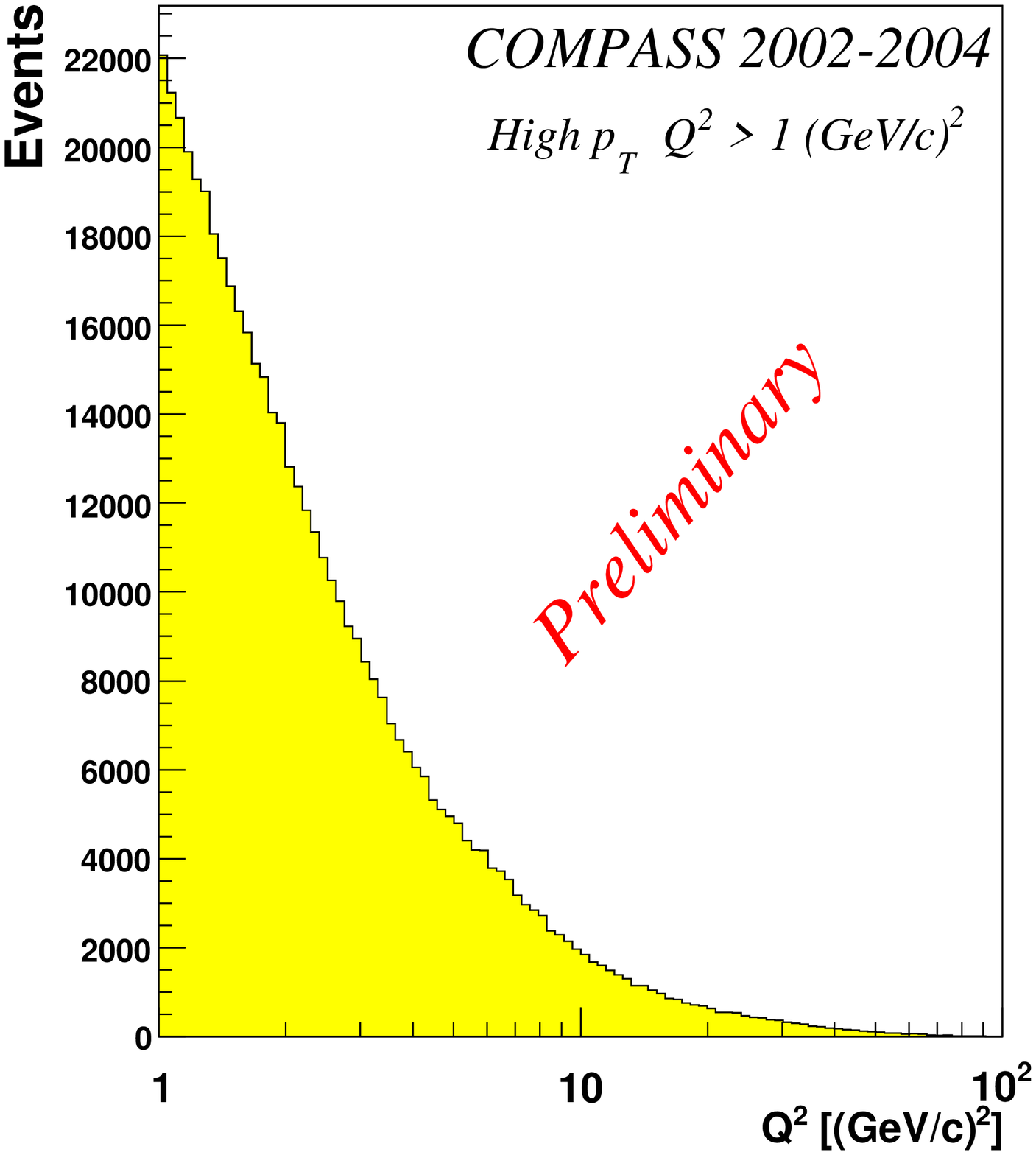}}
  \resizebox{0.32\textwidth}{!}{
    \includegraphics{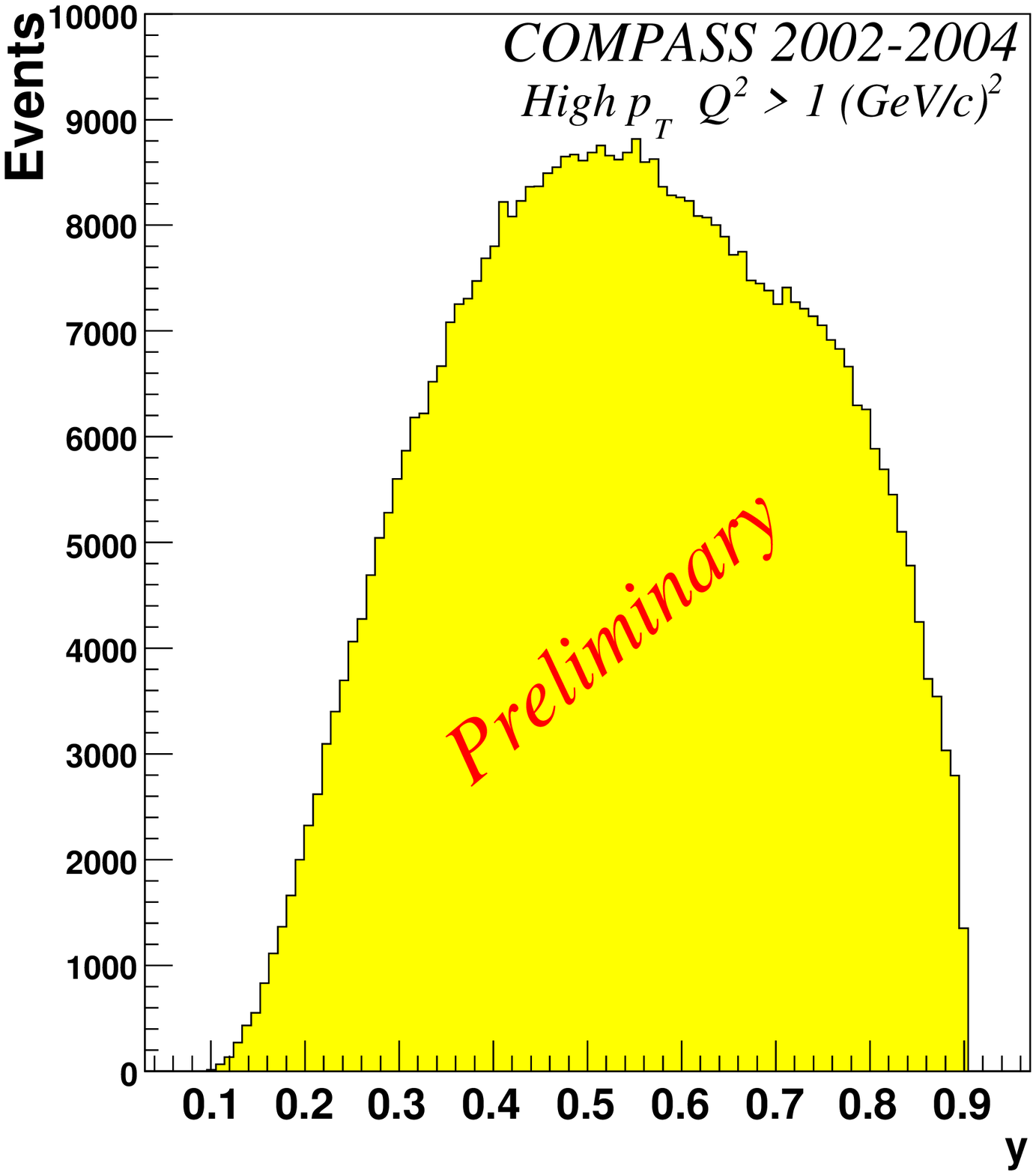}}
  \resizebox{0.32\textwidth}{!}{
    \includegraphics{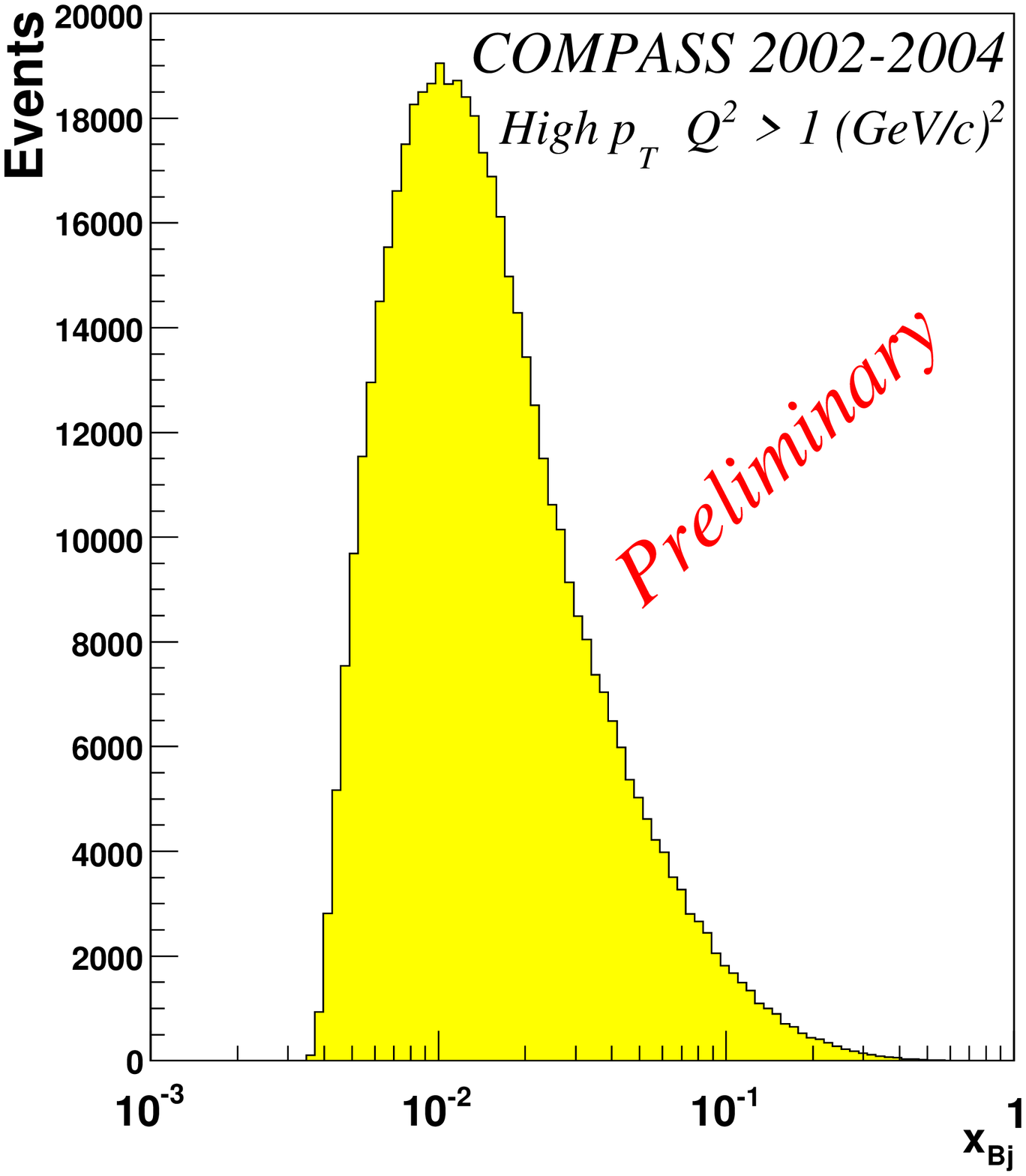}}
\end{center}
\caption{ $Q^2$, $y$ and  $x$ distributions of selected events.}
\label{fig:data:distr}
\end{figure}

\begin{figure}[tbp]
  \begin{center}
  \resizebox{0.6\textwidth}{!}{
    \includegraphics{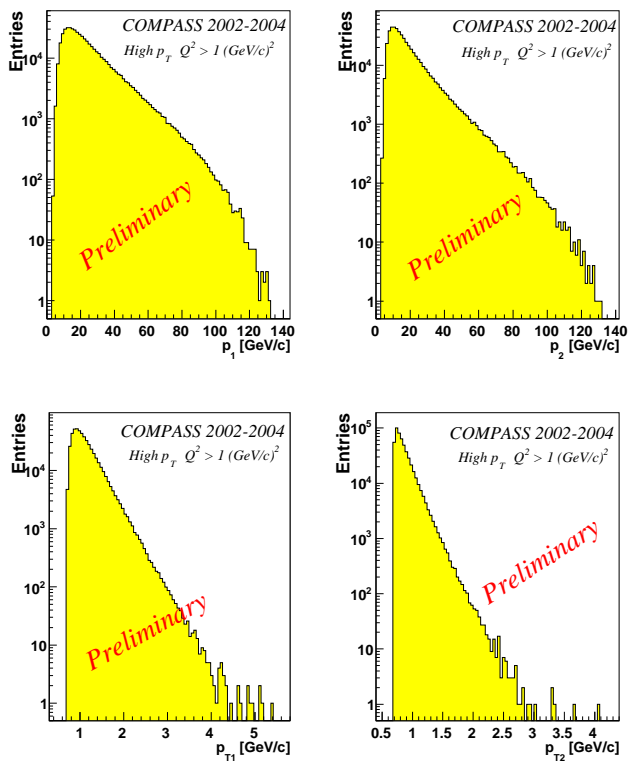}}
\end{center}
\caption{$p$ and $p_T$ distributions for leading and sub-leading hadrons:
  The left column shows the leading hadrons and the right
  column the sub-leading one. In the first row  the momenta are plotted, in the
  second row the transverse momenta are shown.}
  \label{fig:data:p}
\end{figure}

\begin{figure}[tbp]
  \begin{center}
  \resizebox{0.35\textwidth}{!}{
\includegraphics{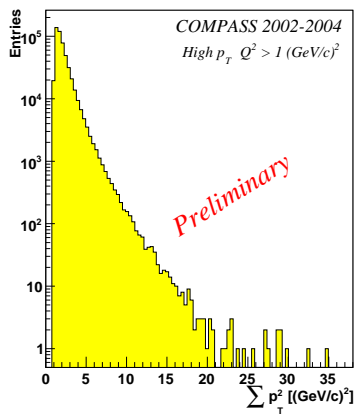}}
\end{center}

\caption{The distribution of the sum of the $p_T^2$ for the leading and sub-leading hadrons.}
  \label{fig:data:zptsum}
\end{figure}

The distributions of the kinematic variables $Q^2$, $y$, $x_{Bj}$
are shown in Fig. \ref{fig:data:distr}. The distributions of
$p$, $p_{T}$, $\sum
p_{T}^{2}$ and $z$ variables are presented in Figs.~\ref{fig:data:p} and \ref{fig:data:zptsum}, for the leading and sub-leading
hadrons.

\section{Monte Carlo simulation}
\label{MC}
A lot of information to be used in
the $\Delta G/G$ calculation is obtained from Monte Carlo (MC) simulation,
therefore this analysis is model dependent. That is the main reason why a
good description of the experimental data by MC is fundamental in this
analysis. Two MC samples were produced to account for the estimation of the
statistical weight: one using the same cuts used in the high $p_T$ event selection
in the previous section (sec. \ref{sec:data}) and another using an inclusive selection based only on the cuts on the DIS kinematic variables ($Q^2$ and $y$). Both MC samples
are restricted to the high $Q^2$ region. The MC production comprises
three steps: First the events are generated, then a spectrometer
simulation program is applied to these events and finally the events
are reconstructed. The spectrometer simulation program based on GEANT3
\cite{geant} was developed. The reconstruction procedure is the same for real and MC events.

For the first step the LEPTO 6.5 \cite{Ingelman:1996mq} 
event generator is used together with a leading order parametrisation of the
unpolarised parton distribution functions with partons generated in a fixed-flavour scheme given by MRST04LO \cite{Martin:2006qz}, with a
good description of $F_2$ in the COMPASS kinematic region. NLO
corrections are simulated by the gluon radiation in the initial and
final state (parton shower ON). The generation is done at two levels:
the simulation of the hard scattering processes and the fragmentation and hadronisation model.

The fragmentation is based on the Lund
string model \cite{Andersson:89} implemented in JETSET \cite{Sjostrand:2000wi}. In this model the
probability that a fraction $z$ of the available energy will be
carried by a newly created hadron is expressed by the Lund symmetric function
$f(z)=z^{-1}(1-z)^a e^{-b m_\perp /z}$, with $m_\perp^2 = m^2 + p_\perp^2$, where
$m$ is the quark mass. To improve the agreement between MC
and data, the parameters ($a$,$b$) were modified from their default values (0.3,
0.58) to (0.6, 0.1), and also the gluon radiation at the initial and
final states were used; i.e. the so-called Parton Shower (PS) ON mode.

\begin{figure}
  \begin{center}
  \resizebox{0.65\textwidth}{!}{
\includegraphics{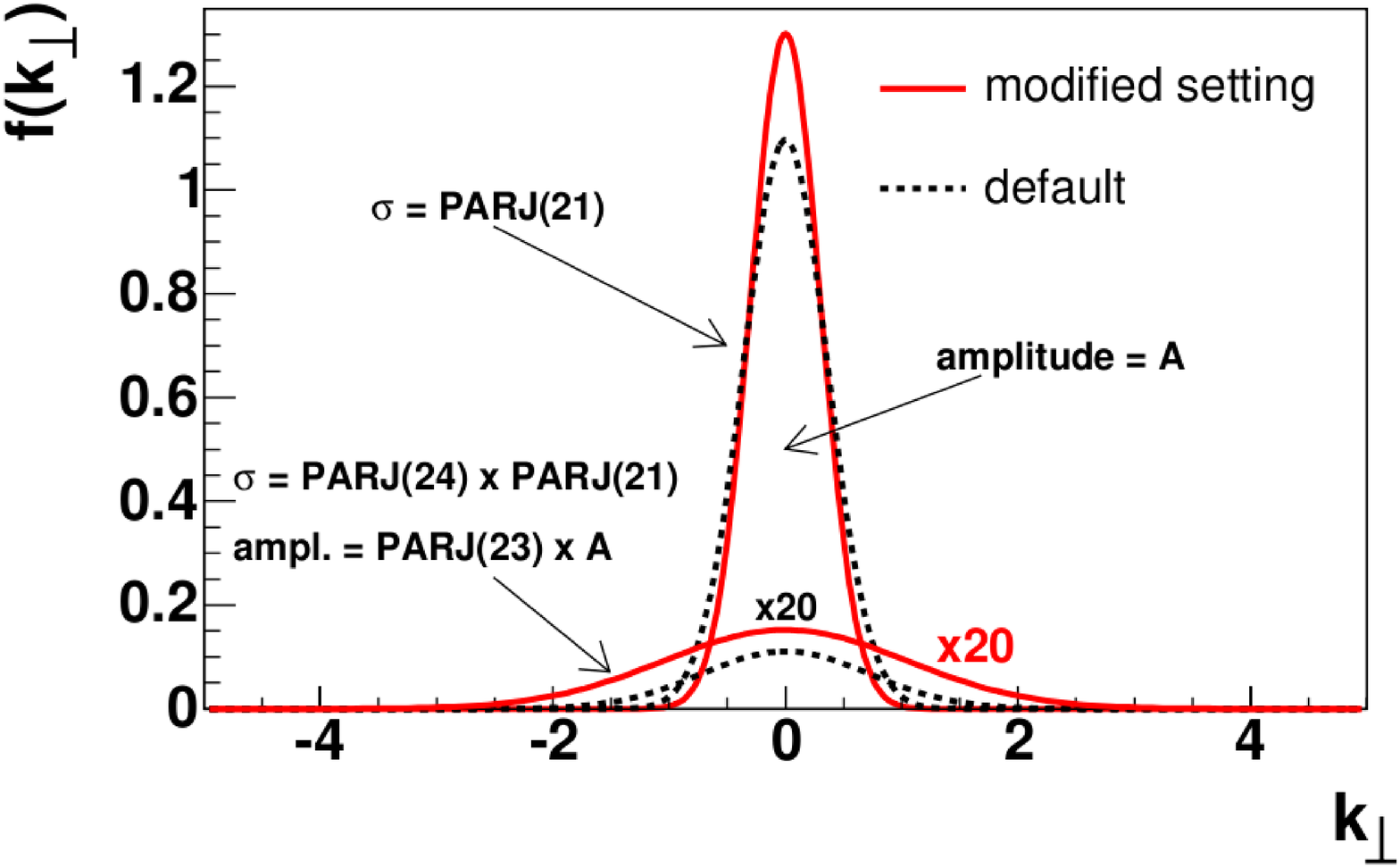}}
\end{center}
\caption{The distribution of $k_T$.}
  \label{fig:kt}
\end{figure}

The transverse momentum of the hadrons, $k_T$, at the fragmentation level is
given by the sum of the $p_\perp$ for $q_i\bar{q_j}$ hadron. Then the
$k_T$ of the newly created hadrons is described by a convolution of two
gaussian distributions; as it is illustrated in
Fig. \ref{fig:kt}, PARJ(21) is the width of the narrower gaussian, PARJ(23) and
PARJ(24) are, respectively, the factors to apply on the amplitude and
on the width for the broader gaussian. The default values of
(PARJ(21), PARJ(23), PARJ(24)) are (0.36 $(\mbox{GeV}/c)^2$, 0.01,
2.0), and were modified to (0.30 $(\mbox{GeV}/c)^2$, 0.02, 3.5).  This set of modifications in these fragmentation parameters it is called the COMPASS tuning.

\begin{figure}[h!]
\centerline{\includegraphics[clip,width=0.33\textwidth]{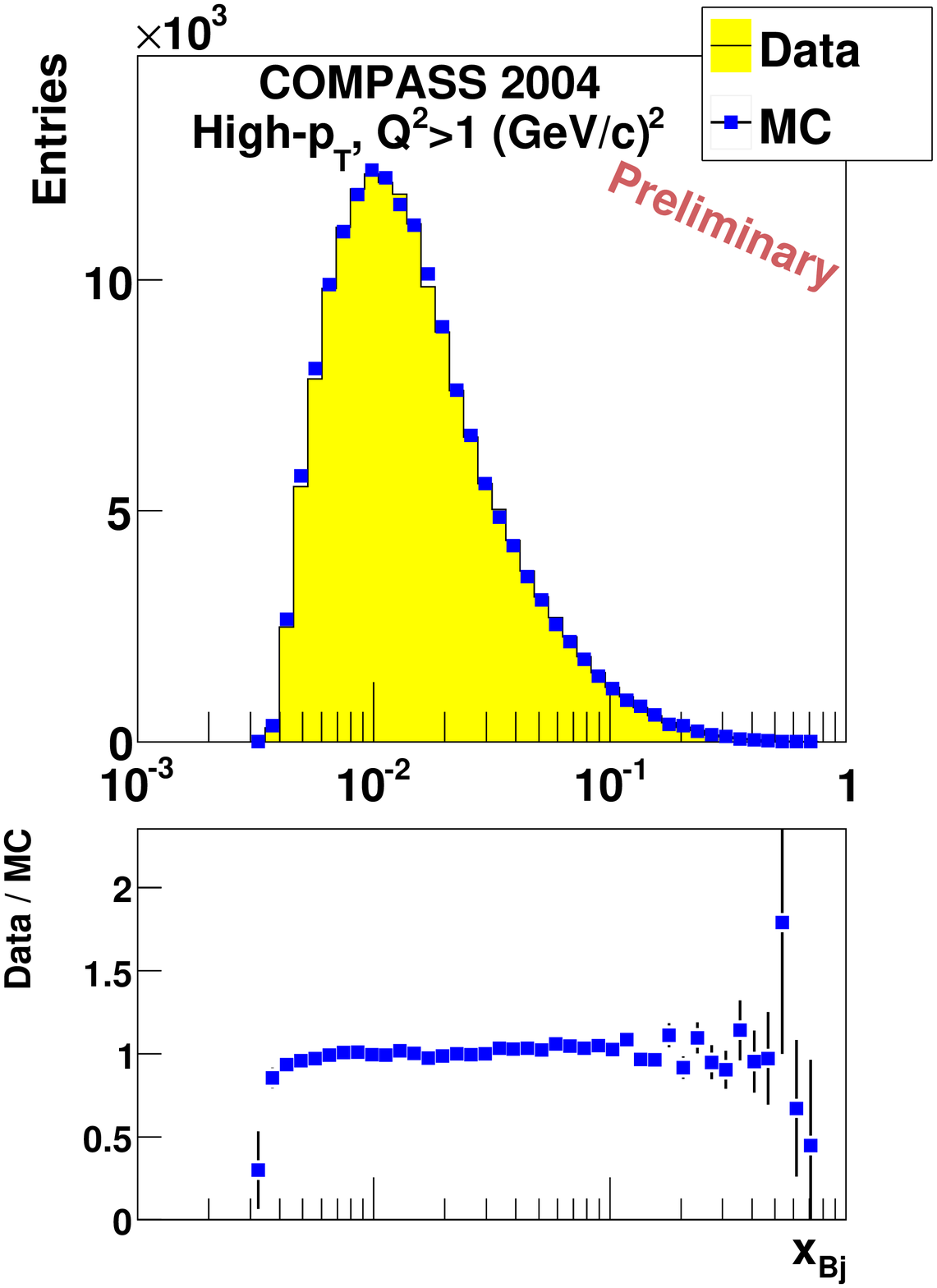}
\includegraphics[clip,width=0.33\textwidth]{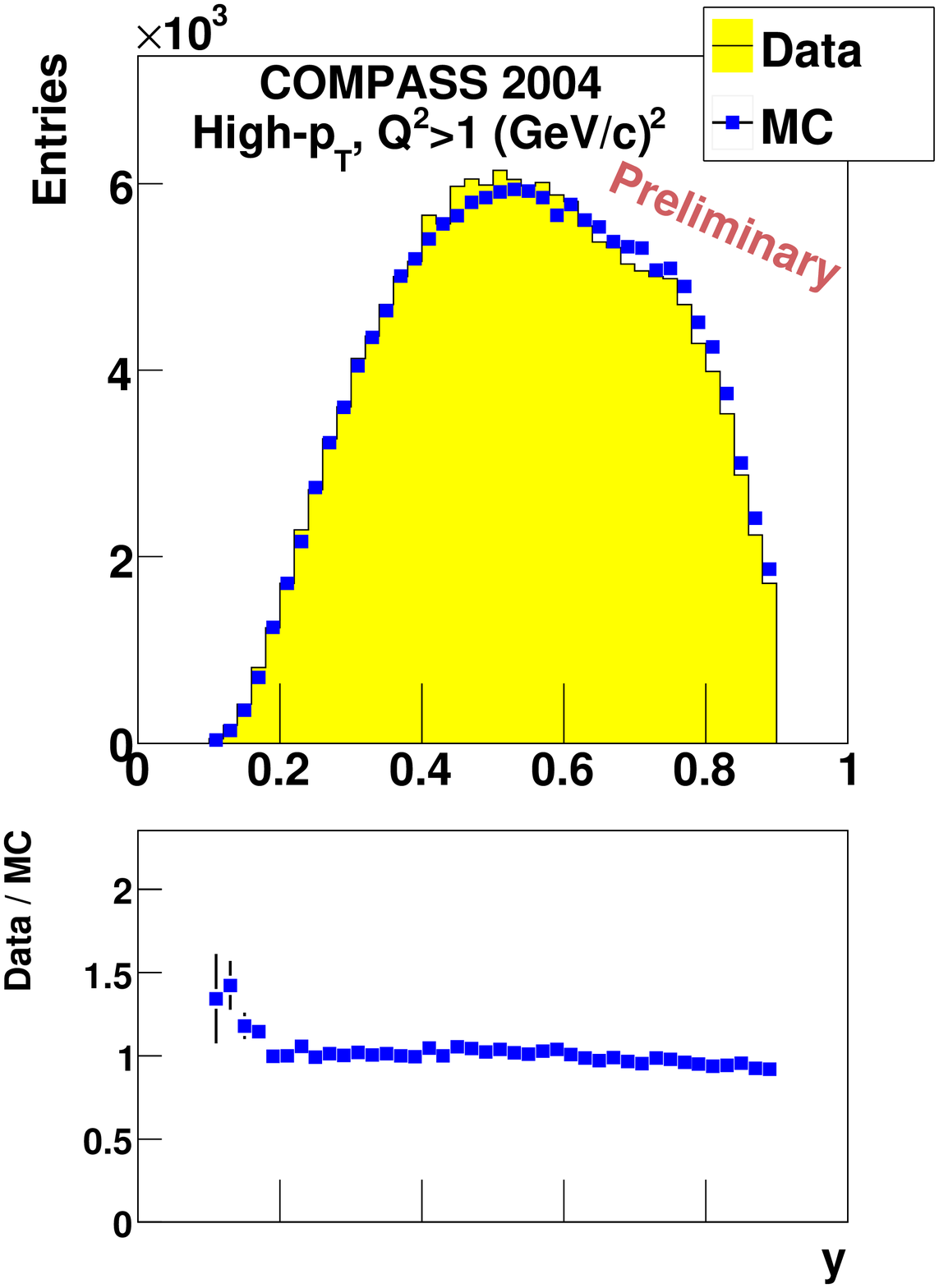}
\includegraphics[clip,width=0.33\textwidth]{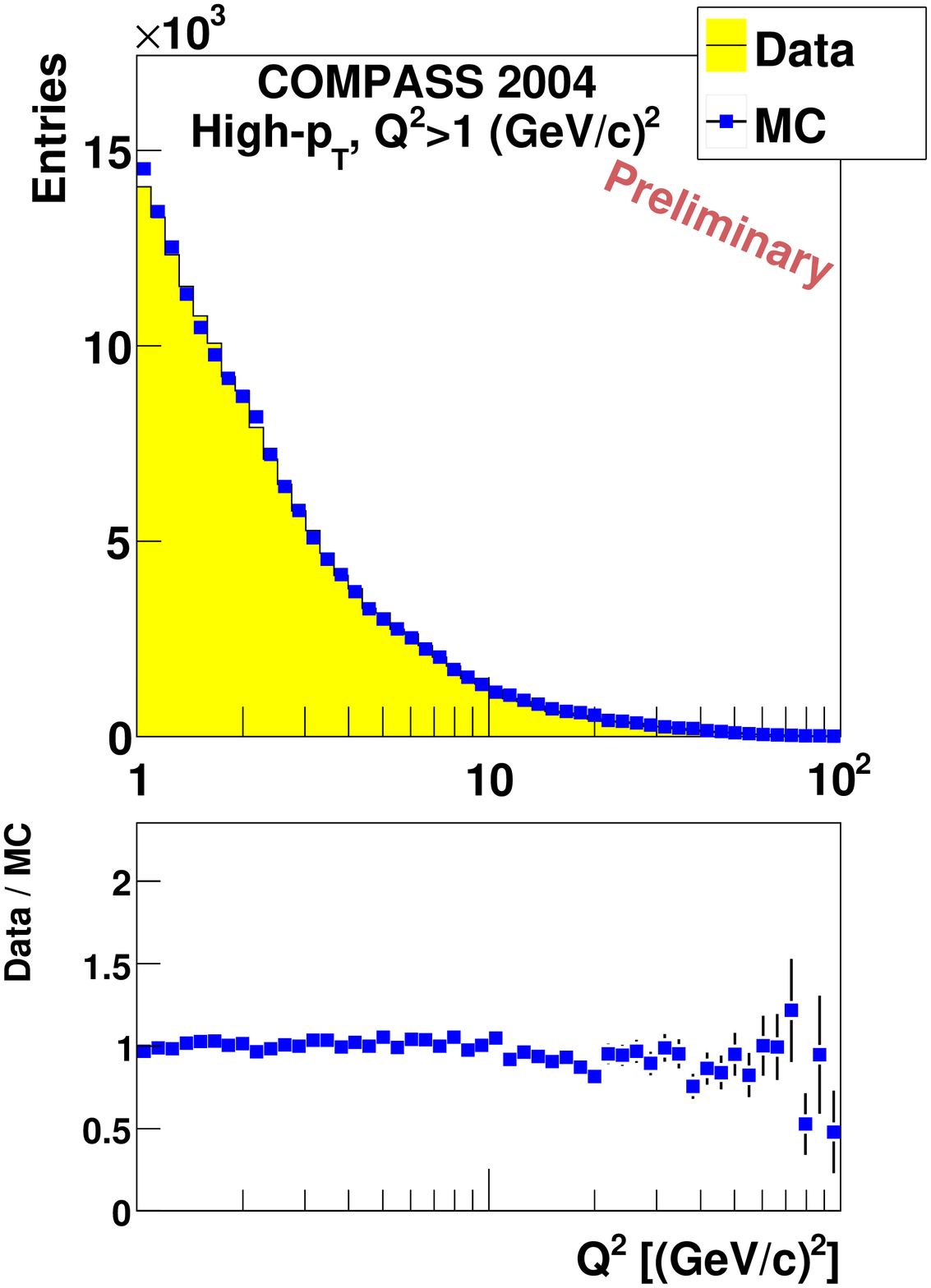}}
\caption{Comparison between data and MC simulation: On plots distributions and ratios Data/MC for inclusive variables
are shown: $x_{Bj},\, Q^{2},\, y$.}
\label{fig:mc:tune_inc}
\end{figure}

\begin{figure}[htb]
\centerline{\includegraphics[clip,width=0.33\textwidth]{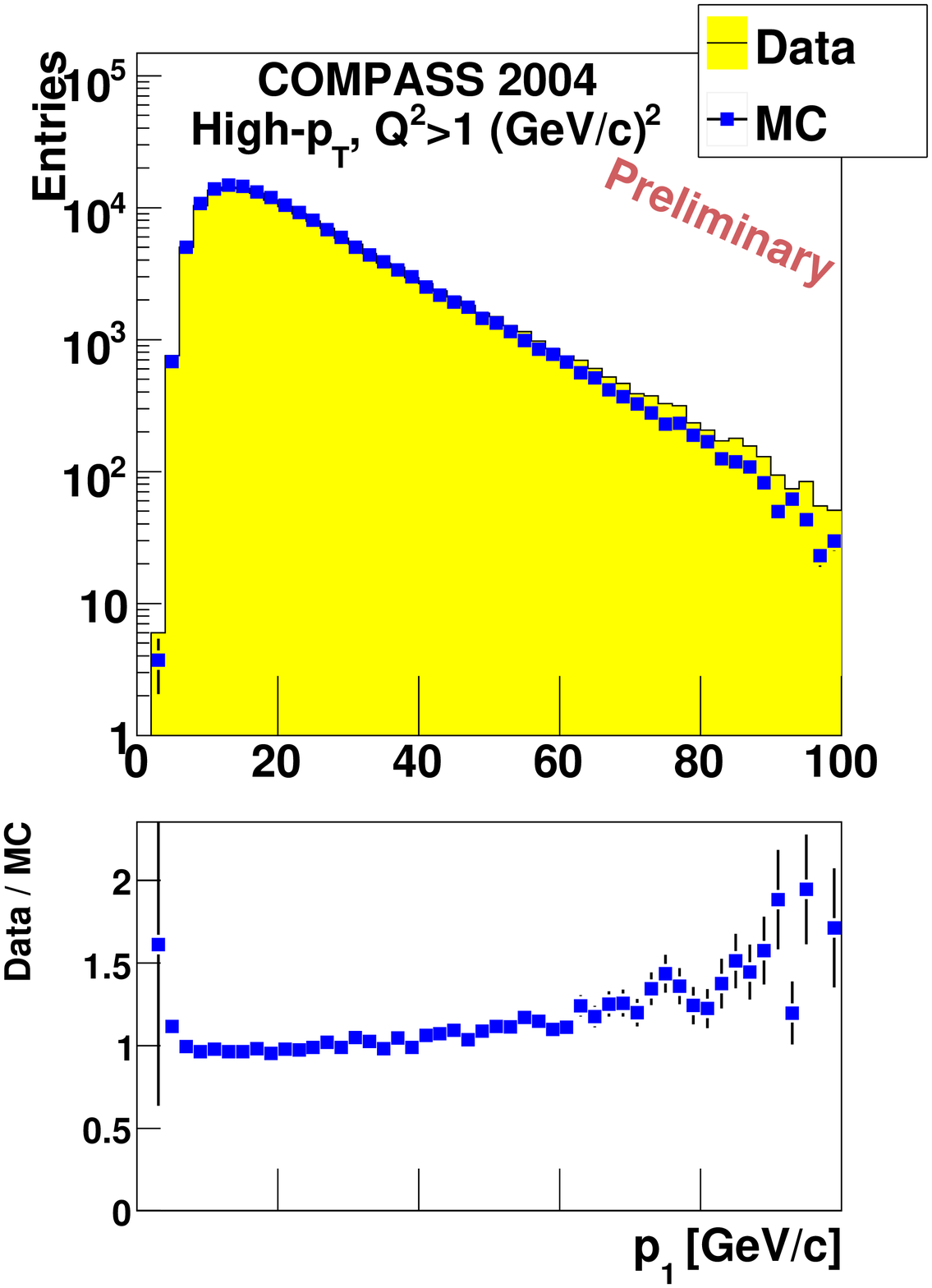}
\includegraphics[clip,width=0.33\textwidth]{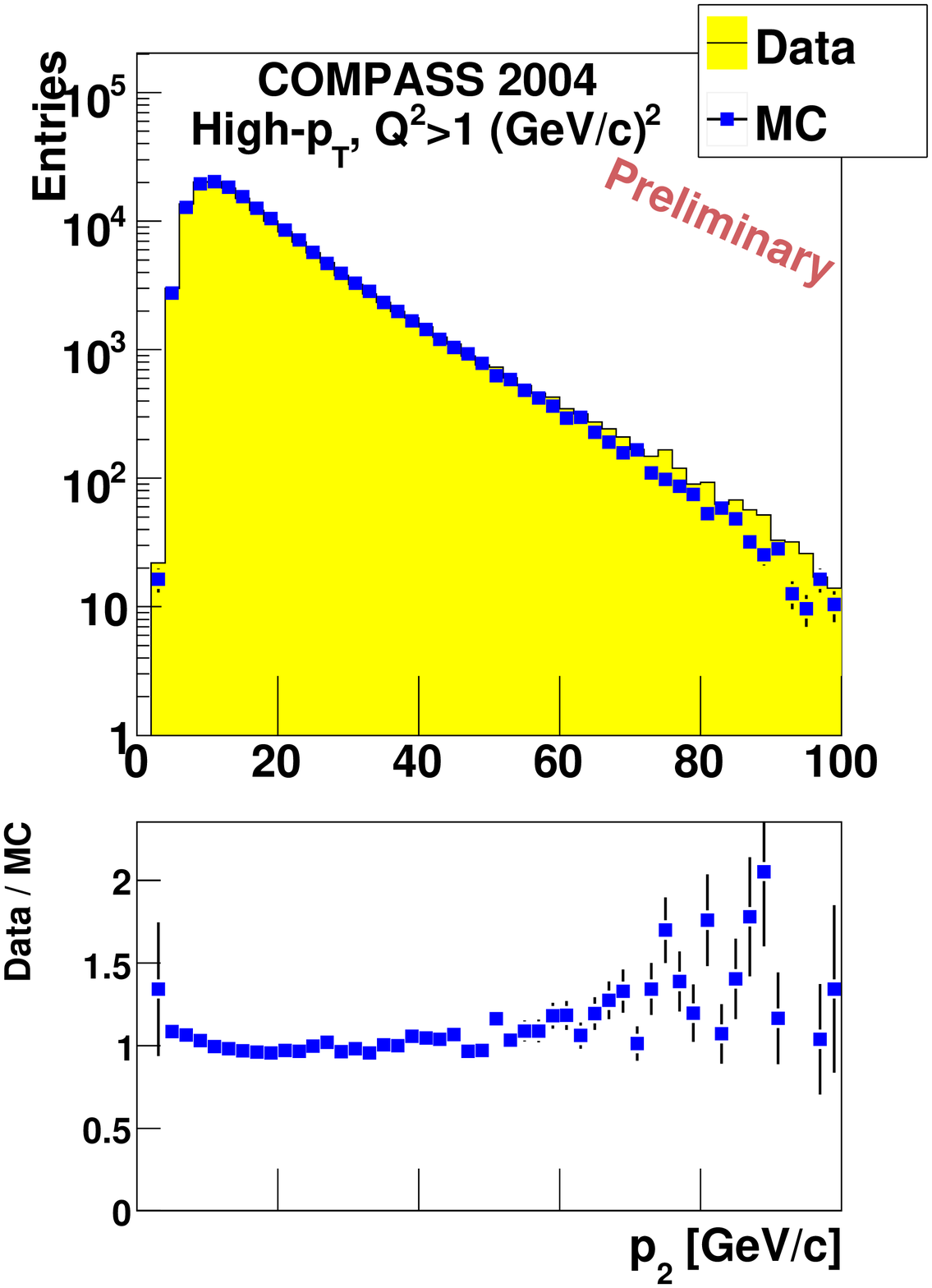}}
\centerline{\includegraphics[clip,width=0.33\textwidth]{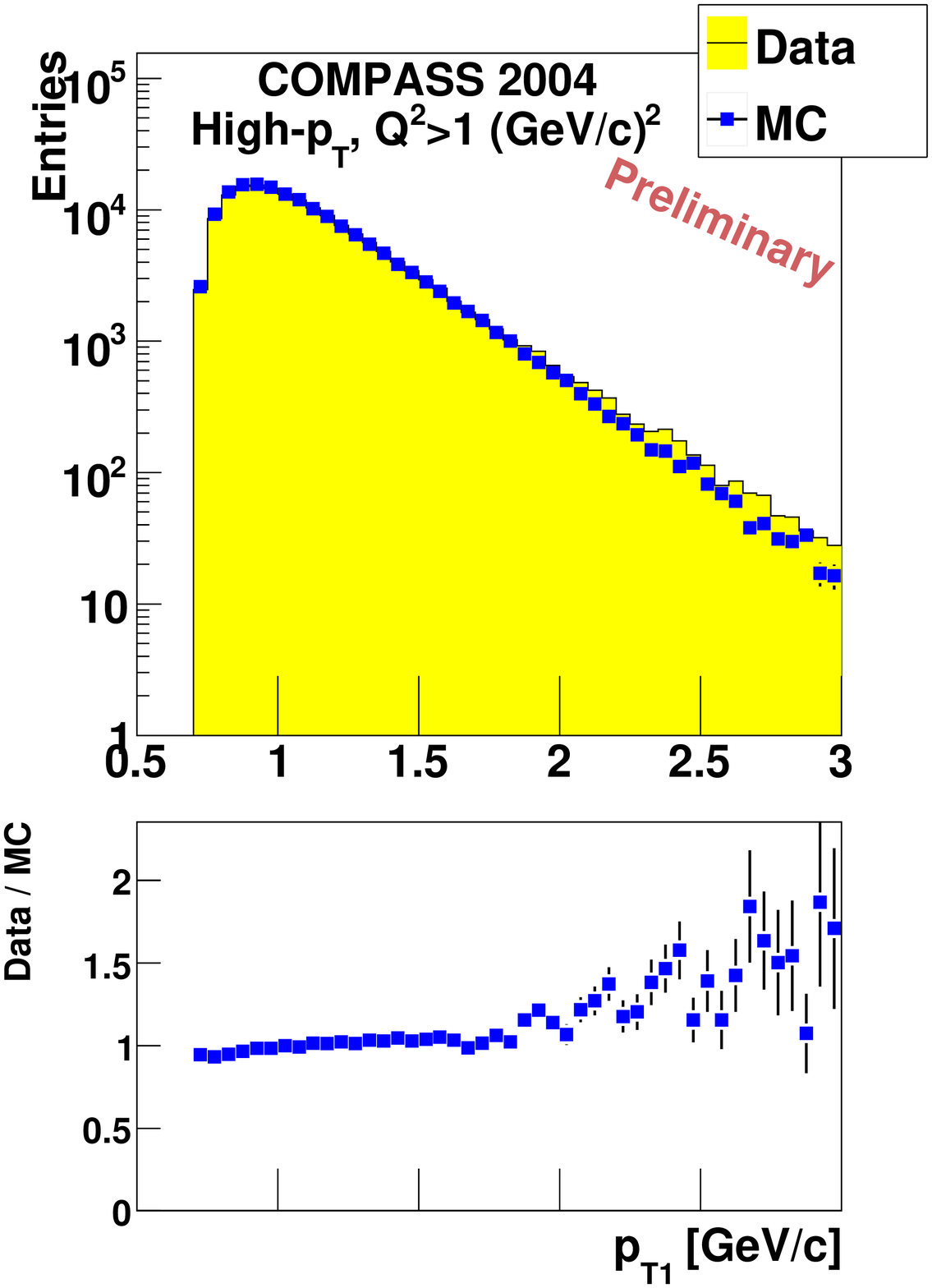}
\includegraphics[clip,width=0.33\textwidth]{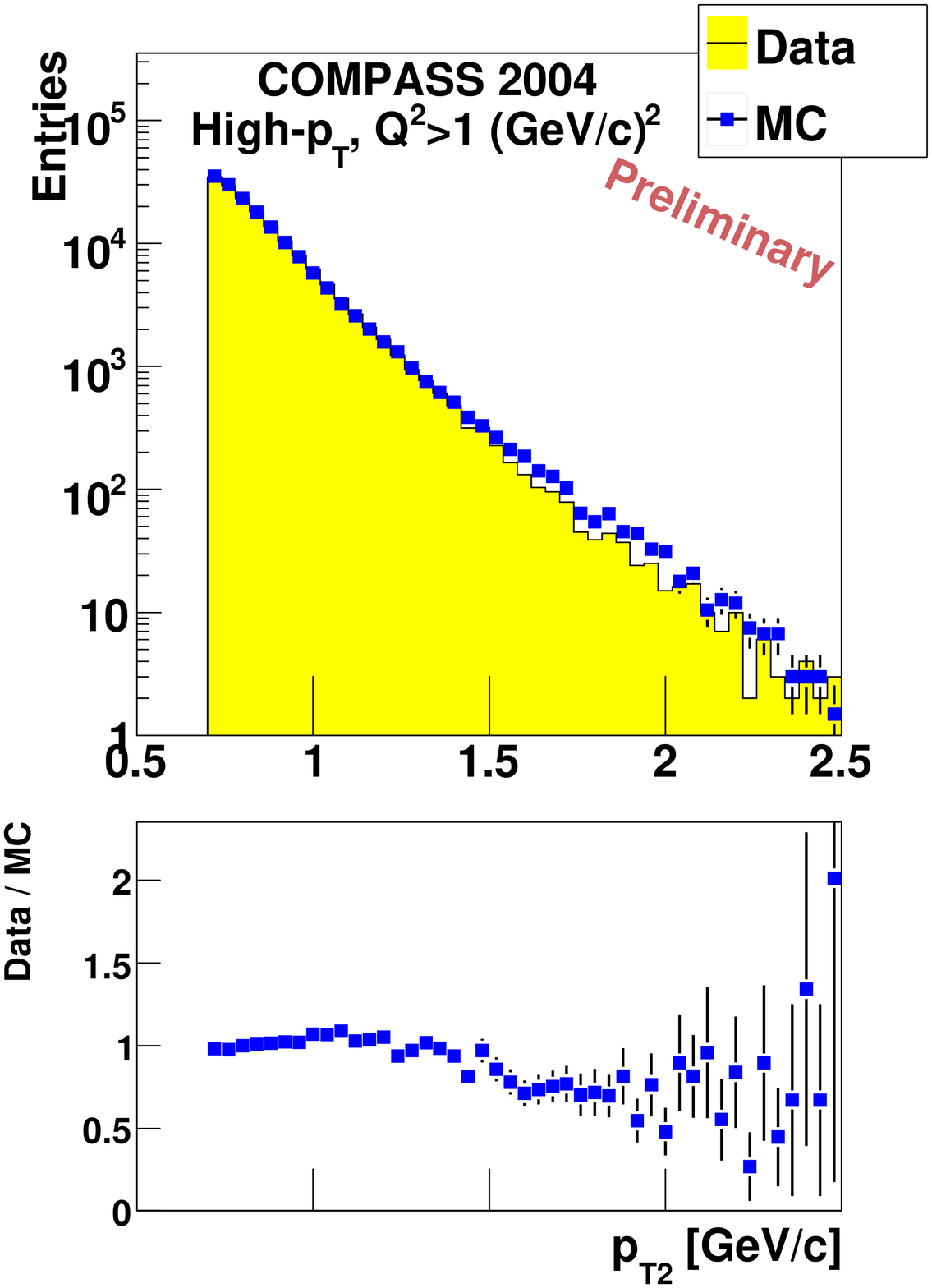}
\includegraphics[clip,width=0.33\textwidth]{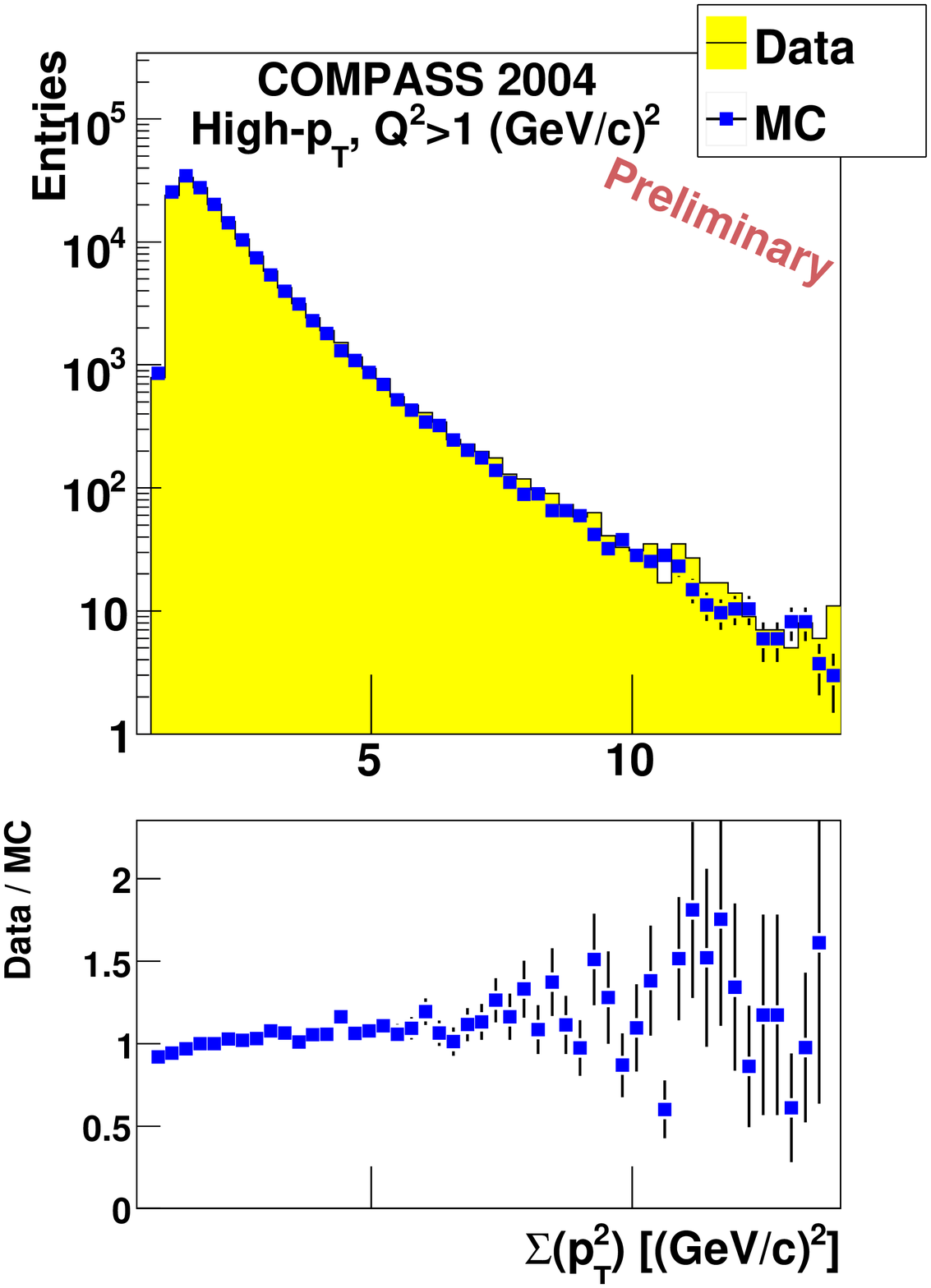}}
\caption{Comparison between data and MC simulation:
 On the plots distributions and ratios Data/MC for hadronic
variables are shown: $p_{1},\, p_{2},\, p_{T1},\, p_{T2}$ and
$\sum(p_{T}^{2}$.}
 \label{fig:mc:tune_had}
\end{figure}

\begin{figure}[htb]
\centerline{\includegraphics[clip,width=0.33\textwidth]{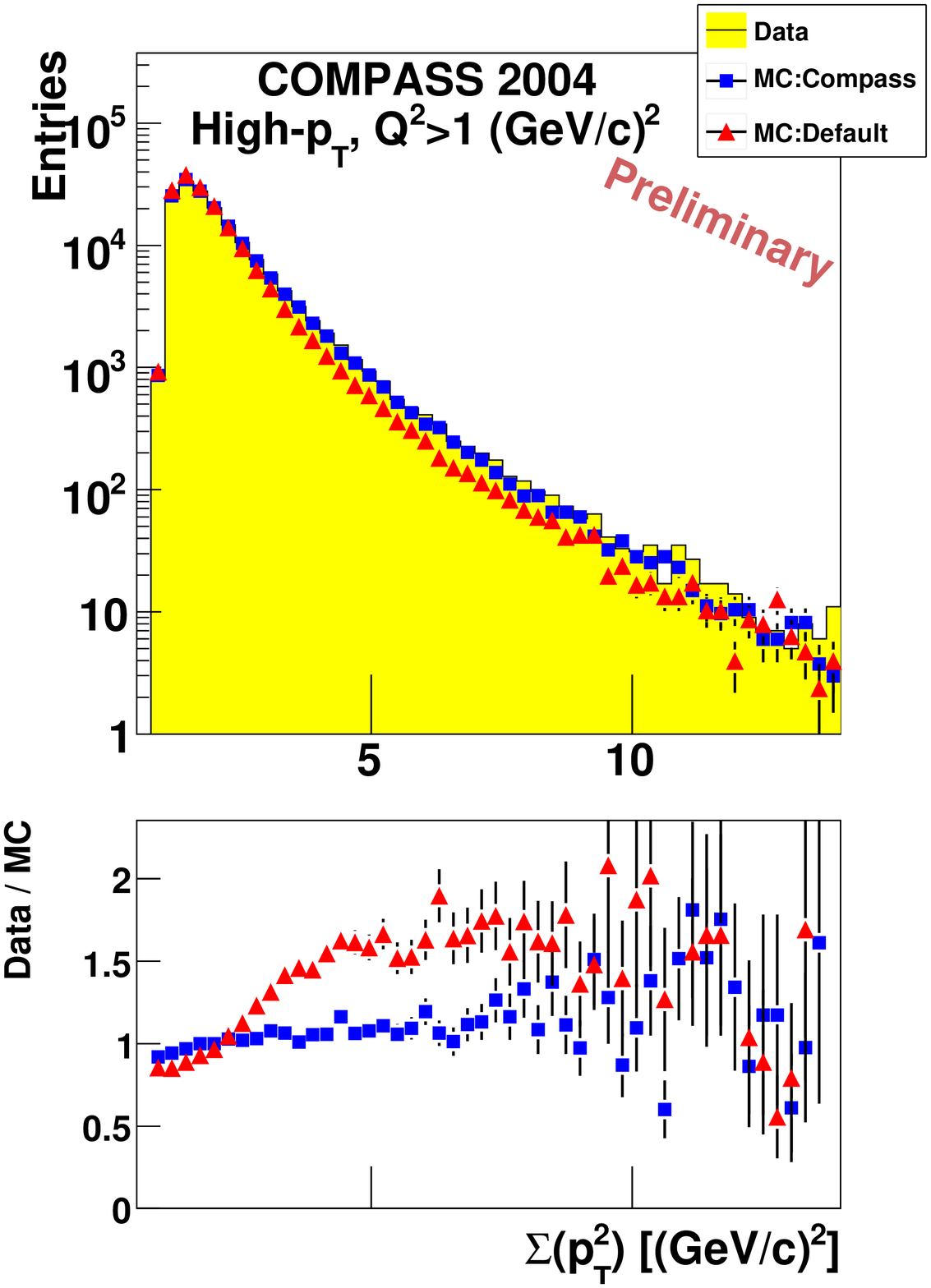}}
\caption{Comparison between COMPASS and default LEPTO MC genetation tunings for $\sum p_{T}^{2}$ variable.}
 \label{fig:mc:tunes}
\end{figure}

The remarkable agreement of the MC simulation with the data is illustrated
in Figs. \ref{fig:mc:tune_inc}-\ref{fig:mc:tunes}. In this analysis the used MC sample has PS on mode.  

The MC--data comparison for different 
the kinematic variables is shown in Fig. \ref{fig:mc:tune_inc}. In
Fig. \ref{fig:mc:tune_had} the hadronic variables, $p$, $p_T$ for the
leading and sub-leading hadrons are shown, together with the sum of
$p_T^2$; i.e. $\sum p_{T1}^2 +p_{T2}^2$. In Fig. \ref{fig:mc:tunes}
the two comparison of the $\sum p_T^2$ variable one using the COMPASS
tuning and another using the default LEPTO tuning
 is also shown the COMPASS tuning describes better our data sample than the LEPTO default one.

\section{The $\Delta G/G$ extracting method}
\label{deltag}
The main goal of this approach is to enhance the PGF
process which accounts for the gluon contribution to the nucleon
spin. In the original idea of the high $p_T$ analysis, the selection was
based on a very tight a set of cuts to suppress LO and QCD
Compton. This situation results in a dramatic loss of statistics.

A new approach was found in which a not so strict set of cuts is
applied, together with a neural network (NN) \cite{Sulej:2007zz} to assign a probability to each event being originated from each of
the three processes.

A parametrisation is created by the neural network using as input 
$x_{Bj}$ and $Q^{2}$, for inclusive sample and while for high $p_{T}$ sample the transverse
and longitudinal momenta of the hadrons, namely $p_{T1}$, $p_{T2}$,
$p_{l1}$, and $p_{l2}$, are used in addition. Using this
parametrisation the values of the event fractions , $R$, the
{\it Bjorken} variable, $x$ and the partonic asymmetry $a_{LL}$ for
each process type are estimeted; these parameters represent the neural
network output.

As the fractions of the three processes sum up to unity, we
need two variables to parameterise them: $o_1$ and $o_2$. While for the
remain output parameters are single. 

The relations between the two neural network outputs $o_{1}$ and $o_{2}$ and
the $R$ fraction are
\begin{itemize}
  \item $R_{\rm PGF} = 1-o_{1}-1/\sqrt{3}\cdot o_{2}$
  \item $R_{\rm QCDC}= o_{1}- 1/\sqrt{3}\cdot o_{2}$
  \item $R_{\rm LO}=  2/ \sqrt{3} \cdot o_{2}$
\end{itemize}

A statistical weight is constructed for each event based on these 
probabilities. In this way we do not need to remove events that most likely do
not came from PGF processes, because the weight will reduce their
contribution in the gluon polarisation calculation. This approach uses
wisely and optimally the statistical power of the available data.

To compute the gluon polarisation in an event-by-event basis the
so-called 2nd order method \cite{jorg} is used. In this analysis several neural networks are used. 

\begin{figure}
  \begin{center}
    \resizebox{0.4\textwidth}{!}{
      \includegraphics{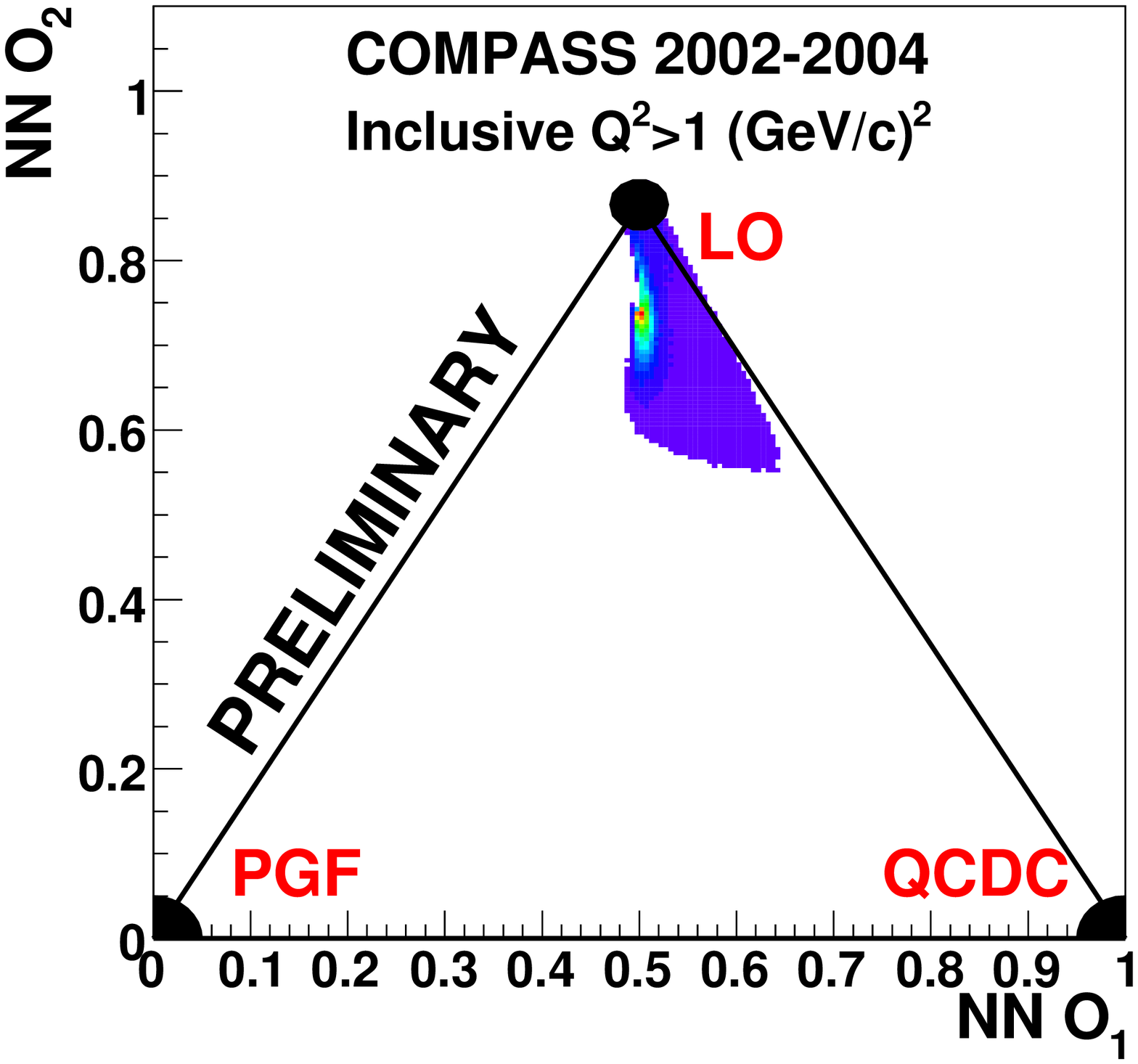}}
    \resizebox{0.4\textwidth}{!}{
      \includegraphics{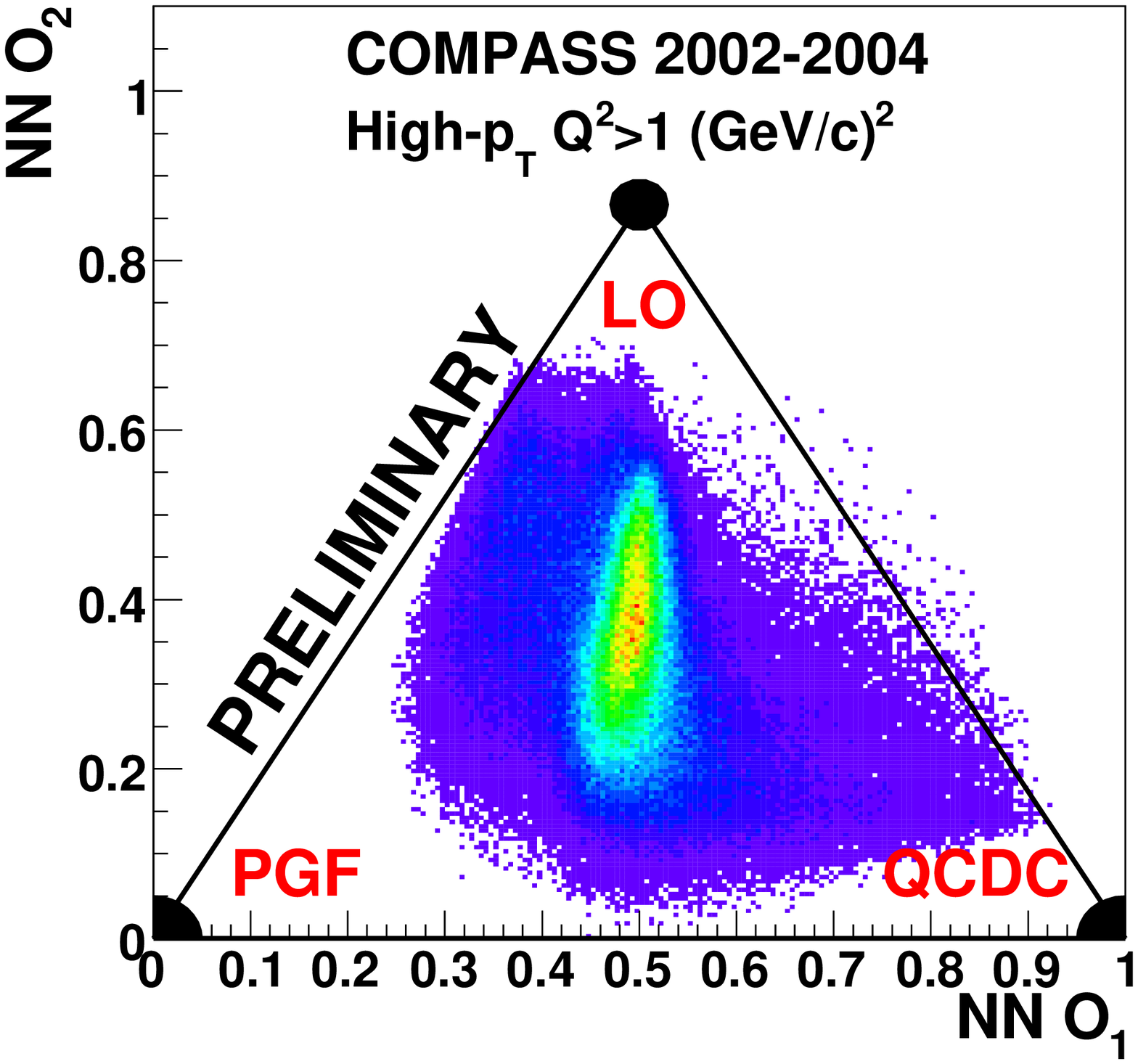}}
  \end{center}
\caption{2-d output of neural network for estimation that the given  event is
PGF, QCDC or LO; (left) for the inclusive
sample and (right) for High $p_{T}$ sample.} \label{fig:form:2dsel}
\end{figure}

The resulting neural network outputs for the fraction are presented in Fig.
\ref{fig:form:2dsel} in a 2-dimensional plot. The triangle
limits the region where all fractions are positive. For the
inclusive sample the average value of $o_2$ is quite large, which
means that the LO process is the dominant one. The situation is
different for the high $p_{T}$ sample, the average outputs are
$\langle o_1 \rangle \approx$ 0.5 and $\langle o_2 \rangle
\approx$ 0.35. Note also that the spread along $o_{2}$ is larger
than along $o_1$.

This means that the neural network is able to detect a region where the contribution of PGF and QCDC is significant compared to LO; although it
can not easily distinguish between the PGF and QCDC processes
themselves.

For the inclusive sample, the $R$ fractions do not depend much
upon $x_{Bj}$ and $Q^{2}$, there is no sharp cut, like e.g.~PGF exists
only for $12.5 \ \mbox{GeV}/c^{2} <Q^{2} < 15 \ \mbox{GeV}/c^{2}$. Therefore the neural network, which gives
the probability that a given event comes from PGF, QCDC or LO,
never produces output close to $(0,0)$, i.e.~$R_{\rm PGF}=1$. Instead
all the inclusive events are confined in a quite limited region of
$o_{1}$ and $o_{2}$. It is worth to note that the inclusive sample contains about 10\% of PGF events.

\begin{figure}
\centerline{\includegraphics[width=12.0cm]{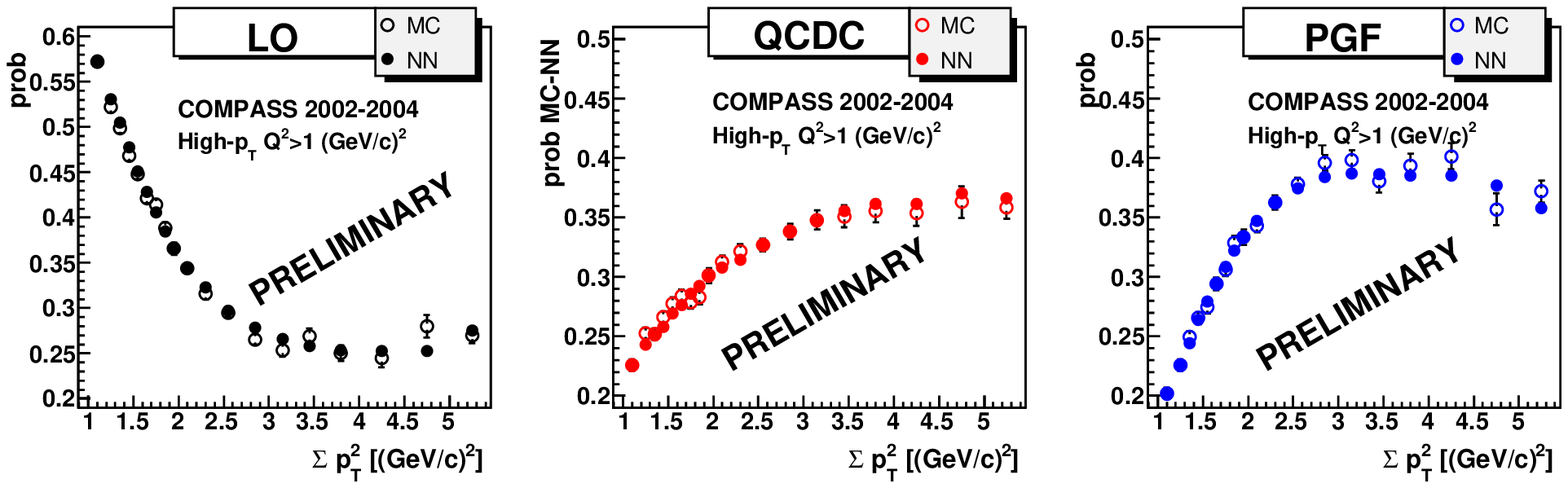}}
\caption{Neural network and MC comparison for $R_{\rm PGF}$, $R_{\rm QCDC}$, $R_{\rm LO}$as a function of $\sum p_T^{2}$.} \label{fig:sys:stabnn_pt}
\end{figure}

To check the stability of the neural network the probability estimated from with the neural network is compared to the fraction of events for each process in MC. As can be seen it in Fig.~\ref{fig:sys:stabnn_pt}, the comparisons are done for each process type separately and in bins of $\sum p_T^2$. The agreement for this comparison is good.

\section{High $p_T$ hadron pair analysis for low $Q^2$ region }
\label{sec:lowq2}

This analysis was published in ref. \cite{Ageev:2005pq}, using a data sample from the years 2002
to 2003. Here in this section, I will describe the same analysis
but using data from 2002 to 2004.  

The reason for splitting all the $Q^2$ range in two
complementary regions is that for low $Q^2$ region the resolved photon contribution is considerably high ($\approx 50 \ \%$). Therefore a more complicated description of the physics than pure QCD in lowest order needs to be included in the MC simulation.

This means that the analysis formalism, which is based on the
processes that are involved in the analysis, is different for both
regions. The data sample in the low $Q^2$ region is 90 \% of all available data.

In the analysis in low $Q^2$ region the selection is essentially the
same as in high $Q^2$ region plus a slightly tighter set of cuts:
$x_F>0.1$, $z > 0.1$, and $\sum p_T^2 >2.5 \ (\mbox{GeV}/c)^2$. 

\begin{figure}
\centerline{\includegraphics[width=10.0cm]{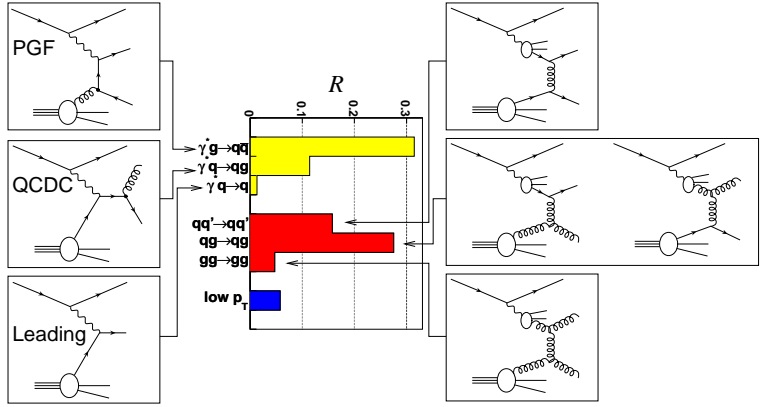}}
\caption{$Q^2$ distribution for the low $Q^2$ region.} \label{fig:lowq2diags}
\end{figure}

This strict set of cuts is used to reduce the physical background
contributions (LO, QCD Compton and resolved photon). The weighting
method used in the high  $Q^2$ analysis, is not applied in this
case. As a consequence not only the physical background is reduced but
also a PGF fraction of events may be reduced too by this strict selection. In Fig. \ref{fig:lowq2diags} the involved processes and their respective ratios are shown.

\begin{figure}[h!]
\centering
\includegraphics[width=0.3\textwidth]{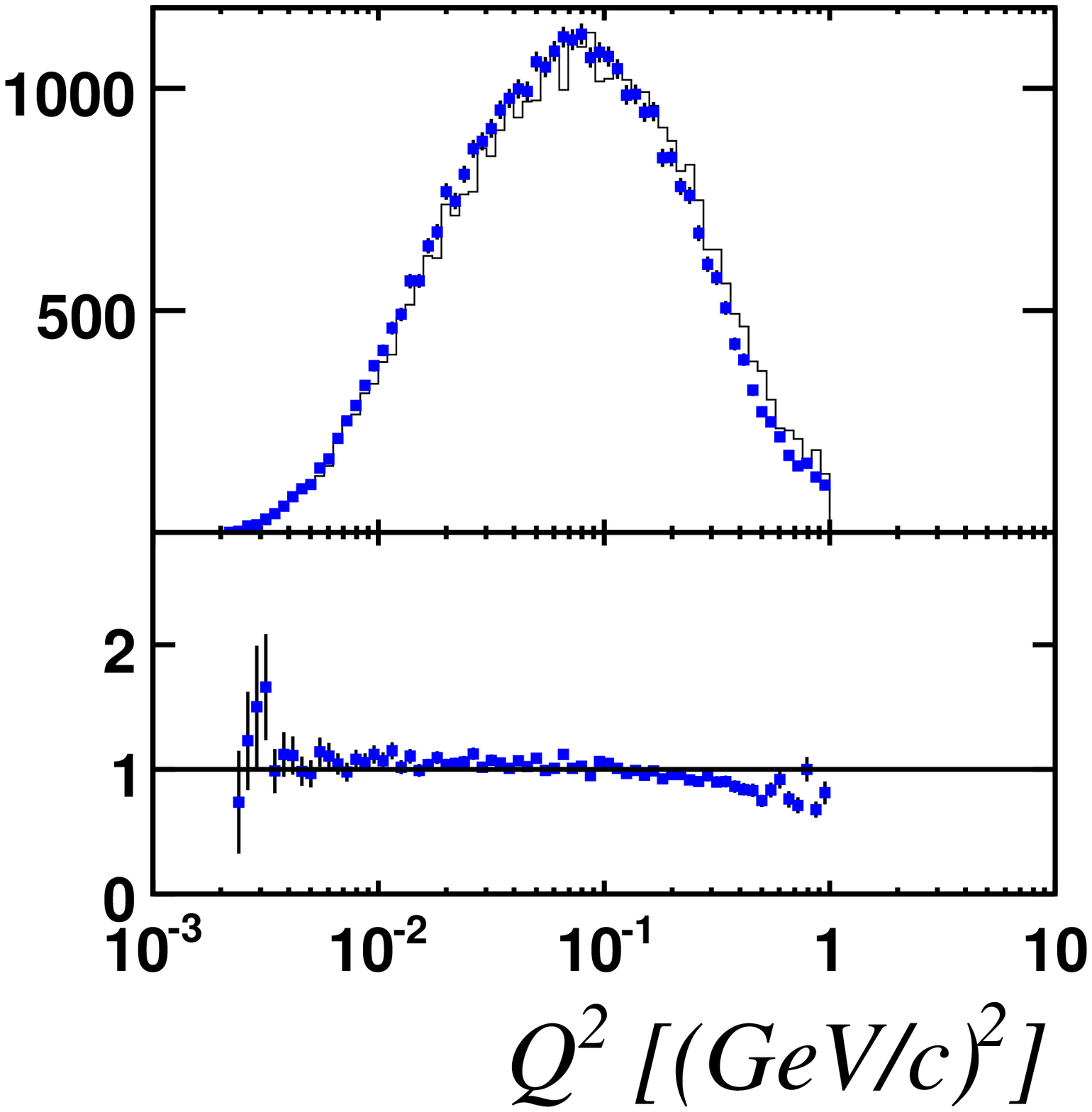}
\includegraphics[width=0.3\textwidth]{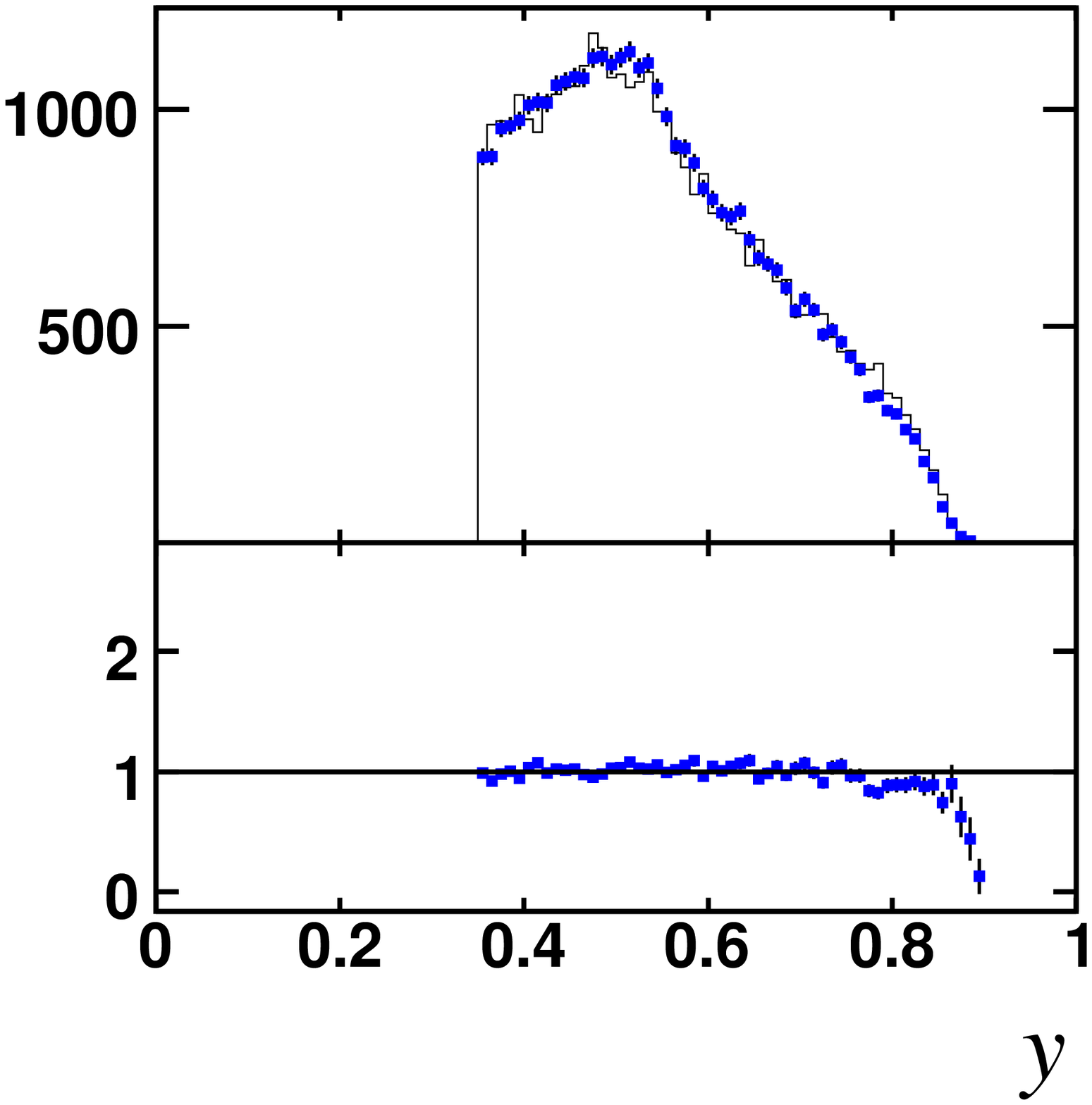}\\
\includegraphics[width=0.3\textwidth]{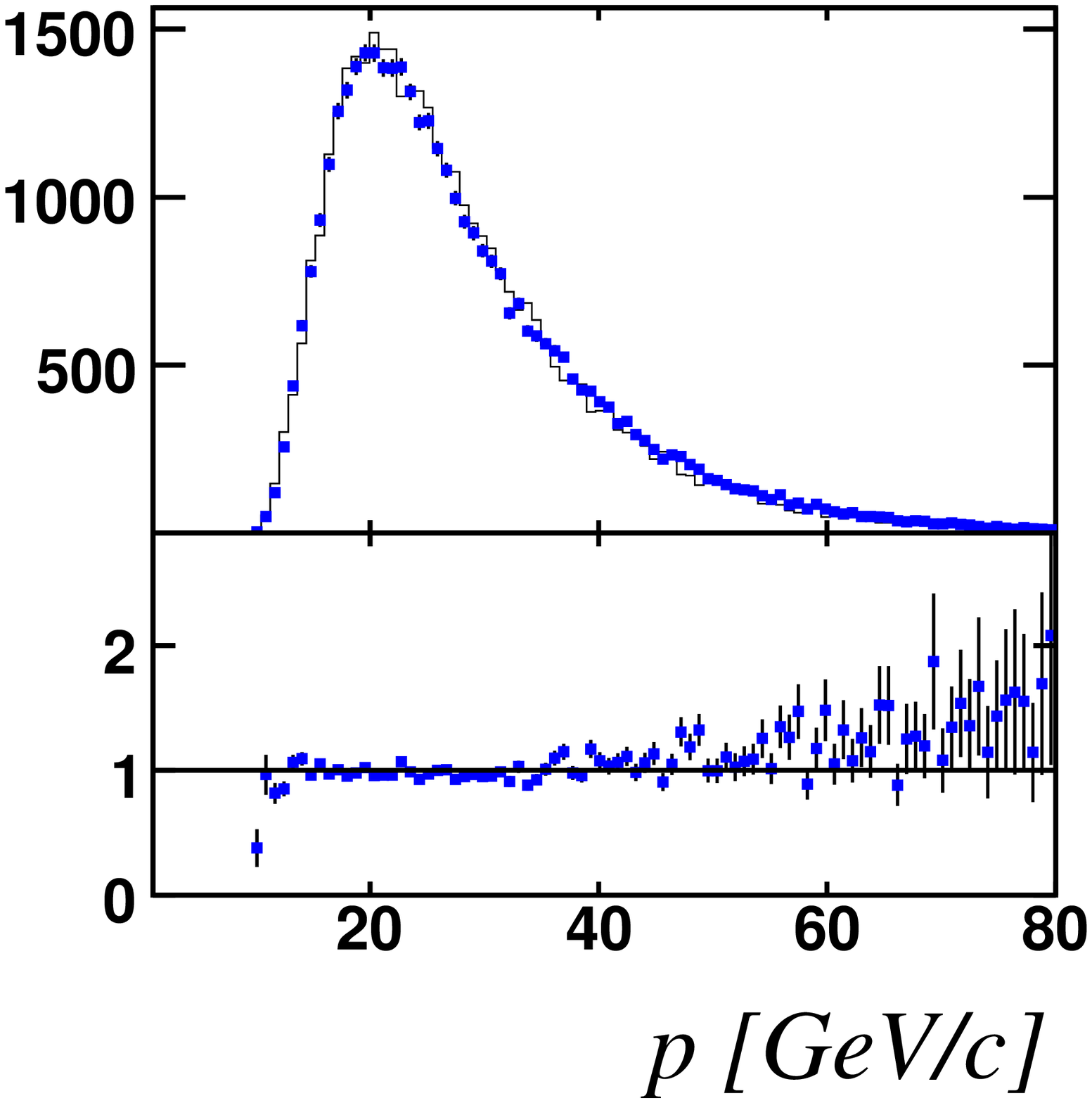}
\includegraphics[width=0.3\textwidth]{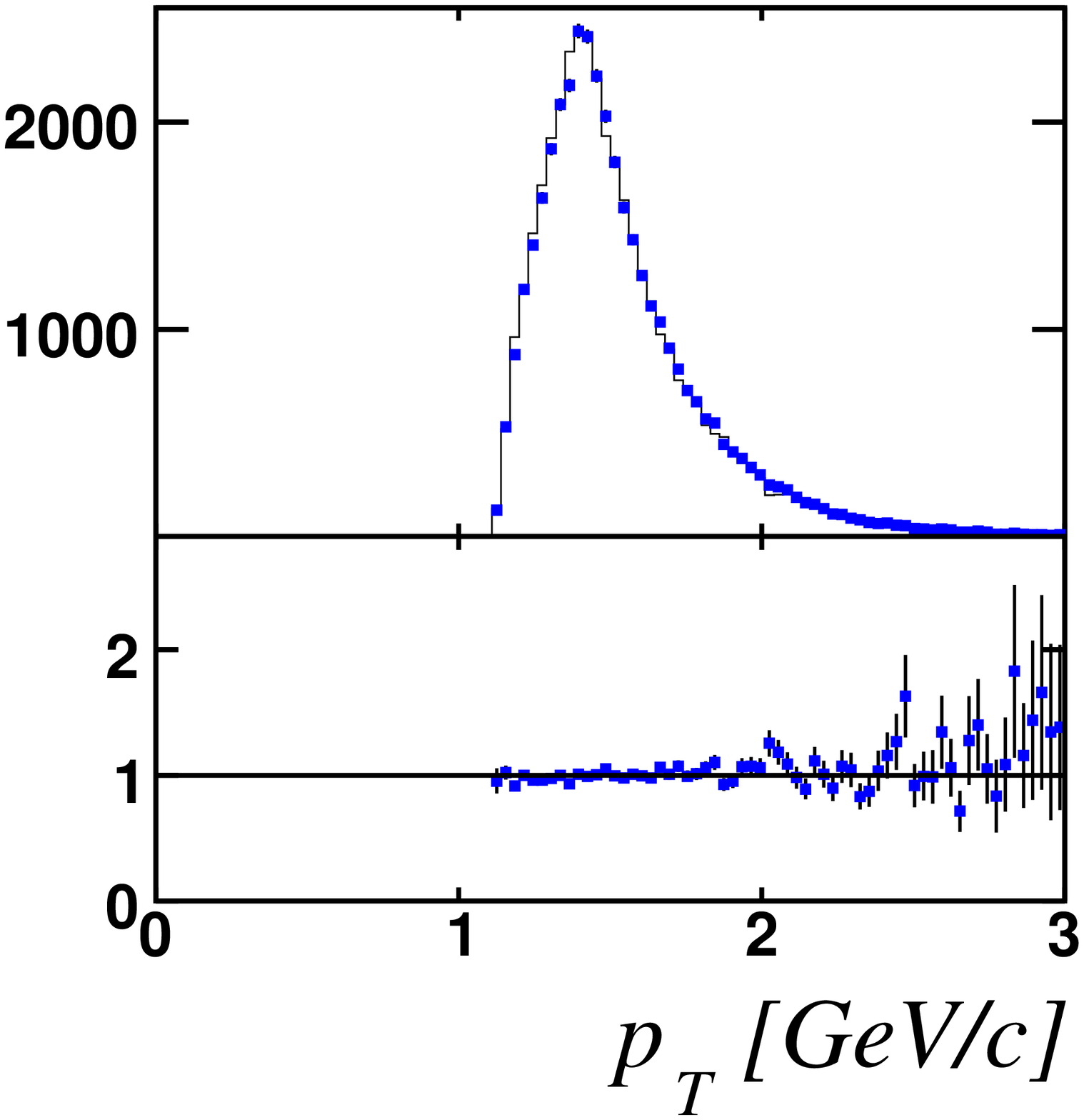}
\caption{Comparison between data and MC for $Q^2$, $y$, and for
the total (transverse) momentum $p$ ($p_{T}$) of the hadron with
highest $p_{T}$. The upper part of each plot shows the  real data
(points) and simulation (line), normalised to the same number of events.
The lower part shows the corresponding data/MC ratio.}
\label{fig:lowq2mcrealcomp}
\end{figure}

The MC simulated and real data samples of high $p_{T}$ events are
compared in Fig.~\ref{fig:lowq2mcrealcomp} for $Q^2$, $y$, and for the total
and transverse momenta of the leading hadron;showing a good
agreement. An equally good agreement is obtained for the sub-leading hadron.

The gluon polarisation in the low $Q^2$ region is extracted using averaged values as shown by this expression:

\begin{eqnarray}
  \nonumber \left \langle \frac{A_{\rm LL}}{D} \right \rangle &=&
  R_{\rm PGF} 
            \left\langle \frac{\hat a_{\rm LL}^{\rm PGF}}{D} \right\rangle
\frac{\Delta G}{G}
  + R_{\rm QCDC} \left \langle \frac{\hat a_{\rm LL}^{\rm QCDC}}{D} A_1 \right \rangle \\
  \nonumber &+& \sum_{f,f^\gamma=u,d,s,\bar u, \bar d, \bar s, G}
R_{ff^\gamma} \left\langle \hat a_{\rm LL}^{ff^\gamma}  \frac{\Delta
f}{f} \frac{\Delta f^\gamma}{f^\gamma} \right\rangle\\
  &+& R_{\rm LO} \times A_{\rm LO} + R_{\rm low-p_T} \times A_{\rm low-p_T}.
\label{eq:allcontribs}
\end{eqnarray}

The hard scale used to compute the gluon polarisation is set by the
$p_T^2$ of the hadrons.

Here, $R_{ff^\gamma}$ is the fraction of events in the whole high $p_{T}$
sample for which a parton $f$ from the nucleon interacts with a parton
$f^\gamma$ from a resolved photon. $A_1$ is the inclusive
virtual photon deuteron asymmetry. $\Delta f/f$ ($\Delta f^\gamma/f^\gamma$) is the polarisation of quarks
or gluons in the deuteron (photon). $R_{\rm low-p_T}$ and $A_{\rm
  low-p_T}$ are respectively the fraction and the asymmetry for events
for which no hard scale can be found, these events are classified in PYTHIA as ``low-$p_T$'' processes. However the asymmetry for this kind
of events is small, as indicated by previous measurements of $A_1$ at low
$Q^2$~\cite{Adeva:1999pa}. Moreover, the leading and low-$p_{T}$
processes together account for only 7\% of the high $p_{T}$ sample.
For these two reasons, we neglected their contributions.

The PGF analysing power $\hat a_{\rm LL}^{\rm
PGF}$ is calculated using the leading order expressions for the
polarised and unpolarised partonic cross sections and the parton
kinematics for each PGF event in the high $p_{T}$ MC
sample.

The contribution of QCD Compton events to the high $p_{T}$ asymmetry
is evaluated from a parametrisation of the virtual photon deuteron
asymmetry $A_1$ based on a fit to the world
data~\cite{Adeva:1998vv,Abe:1998wq}. This asymmetry is calculated for
each event at the momentum fraction $x_{\rm q}$ of the quark, known in
the simulation.

The parton from a resolved photon interacts either with a quark or a
gluon from the nucleon. In the latter case, the process is sensitive to
the gluon polarisation $\Delta G/G$. The analysing powers $\hat a_{\rm
LL}^{ff^\gamma}$ are calculated in pQCD at leading
order~\cite{Bourrely:1987gp}. The polarisations of the $u$, $d$ and $s$
quarks in the  deuteron $\Delta f/f$ are calculated using 
the polarised parton distribution functions from Ref.~\cite{Gluck:2000dy} (GRSV2000) and
the unpolarised parton distribution functions from
Ref.~\cite{Gluck:1998xa} (GRV98, also used as an input for PYTHIA), all
at leading order. The polarisations of quarks and gluons in the virtual
photon $\Delta f^\gamma/f^{\gamma}$ are unknown because  the polarised
parton distribution functions of the virtual photon have not yet been measured. Nevertheless,
theoretical considerations provide a minimum and a maximum value for
each $\Delta f^\gamma$,
in the so-called  minimal and  maximal scenarios~\cite{Gluck:2001rn}.

\section{Results}
\label{res}
The preliminary measurements of the gluon polarisation in low and high $Q^2$
regions, using data from the years 2002 to 2004, are:

\begin{eqnarray*}
  \left(\Delta G/G \right)_{{\rm low} \ Q^2} &=& 0.02 \pm 0.06_{(stat.)} \pm
  0.06_{(syst.)} \qquad {\rm with} \quad x_G= 0.09^{+0.07}_{-0.04}\\
  \left(\Delta G/G \right)_{{\rm high} \ Q^2} &=& 0.08 \pm 0.10_{(stat.)} \pm 0.05_{(syst.)} \qquad {\rm with} \quad x_G= 0.08^{+0.04}_{-0.03} 
\end{eqnarray*}

The average of the hard scale, $\mu^2$, for low and high $Q^2$ is 3
$(\mbox{GeV}/c)^2$. $x_G$ is the momentum fraction carried by
the probed gluons and it is determined by the $x_{Bj}$ distribution
for the PGF processes, obtained from the MC parton kinematics. 

The published result of this measurement for low $Q^2$ using data from
2002 to 2003 can be found in this ref. \cite{Ageev:2005pq}. while for
the low $Q^2$ measurement obtained using data from 2002 to 2004 was
presented in several conferences, therefore this result is presented
for comparison with the high $Q^2$ one.

Concerning the systematic error related to the gluon polarisation in the high $Q^2$ region
several studies were performed to estimate its value.

For the neural network contribution to the systematic error, several neural networks with different fixed number of neurons in their internal structure were
trained on the same MC sample, and for each $\Delta G/G$ was
extracted. 

For the MC contribution several tests were
performed with different fragmentation and initial and final gluon radiation.

The relative error of $\delta P_{b}$, $\delta P_{t}$ is taken as
5\% and 2\% for relative error of the dilution factor, $\delta f/f$. It is assumed that the systematic
error of $\delta(\Delta G/G)_{fP_{b}P_{t}}$ is proportional to the
errors given above. 

The false asymmetries are extracted using
several constraints in the spectrometer with the purpose to assess its
stability. Two main issues have a significant contribution
to the false asymmetries. The first one is related to the asymmetry computed
using acquired data from positive and negative microwave
configurations separately. The second issue is related to the
asymmetry calculated form different regions of the spectrometer (top,
bottom, left and right regions).

A world data fit, with all $Q^{2}$ range, \cite{compassrho} is used to
parametrise $A_{1}^{d}$, to extract the preliminary value of $\Delta G/G$. Four different parameterisations of $A_{1}^{d}$ are used to estimate the
associated systematics.

The impact of the $x'_{C}$ factor in the eq. (\ref{eq:form:gluon})
was estimated in a two tests. In the first one  $x'_{C}$ was
assumed to be proportional to $x_{C}$. In the second $x'_{C}$ was approximated by using $x_{C}$
instead of $x_{Bj}$ as an input parameter for the neural network which estimates
$x_{C}$.

The
main contributions to this uncertainty are summarised in Table
\ref{tab:sys:sumsys} together with their estimated values.

\begin{table}[htb]
\begin{center}
\begin{tabular}{|c|c|}
\hline
 $\delta( \Delta G/G)_{NN}   $          &   0.006  \\ \hline
 $\delta( \Delta G/G)_{MC}   $          &   0.040  \\ \hline
 $\delta( \Delta G/G)_{f,P_{b},P_{t}} $ &   0.006  \\ \hline
 $\delta( \Delta G/G)_{false}         $ &   0.011  \\ \hline
 $\delta( \Delta G/G)_{A1^{d}}        $ &   0.008  \\ \hline
 $\delta(\Delta G/G)_{formula}        $ &   0.013  \\ \hline \hline
  TOTAL  &    0.045   \\ \hline
\end{tabular}
\caption{Summary of the major systematic contributions }
\label{tab:sys:sumsys}
\end{center}
\end{table}

\section{Conclusions and outlook}

The preliminary values of the gluon polarisation for low and high
$Q^2$ regions were presented. The gluons were probed at an average
scale $\langle \mu^2 \rangle \approx 3 \ (\mbox{GeV/}c)^2$.

\begin{figure}[h!]
\centering
\includegraphics[width=0.6\textwidth]{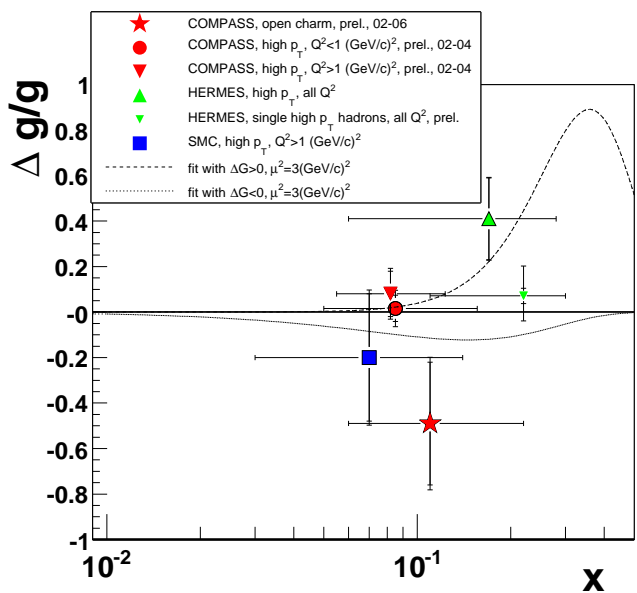}
\caption{Comparison of $\Delta G/G$ measurements from COMPASS \cite{comp.del_sigma}, SMC
  \cite{Adeva:2004dh}, and  HERMES \cite{Airapetian:1999ib}. The two
  curves correspond to the parametrisation from the NLO QCD analysis
  in the $\overline{MS}$ scheme with scale at 3 $(\mbox{GeV/}c)^2$}
\label{fig:dgg}
\end{figure}

Fig. \ref{fig:dgg} shows these new values of $\Delta G/G$ together with the
preliminary value from the open charm analysis \cite{comp.del_sigma} and  the measurements
from SMC collaboration, from the high $p_T$ analysis for
the $Q^2 >1 \ (\mbox{GeV/}c)^2$ region \cite{Adeva:2004dh} and also the measurements from
HERMES collaboration; for single hadrons and high $p_T$ hadrons pairs
analyses \cite{Airapetian:1999ib}. The curves in Fig. \ref{fig:dgg} are the parametrisation of $\Delta
G/G (x)$ using a NLO QCD analysis in the $\overline{MS}$ scheme with a
renormalisation scale $\langle \mu^2 \rangle = 3 \ (\mbox{GeV/}c)^2$. The curve
with the dashed line is the QCD fit assuming that $\Delta G > 0$, the
dotted line is the  QCD fit assuming $\Delta G < 0$. It is seen that
both high $p_T$ values; for high and low $Q^2$ analyses are in
compatible with each other and also, within their $x_G$ region, in
agreement with the NLO QCD fits. These both measurement show that
gluon contribution to the spin for $x_G \approx 0.1$ is compatible
with zero.   

\begin{thebibliography}{}


\bibitem{Alguard:1976bm}
  M.~J.~Alguard {\it et al.},
  Phys.\ Rev.\ Lett.\  {\bf 37} (1976) 1261.

\bibitem{Baum:1983ha}
  G.~Baum {\it et al.},
  Phys.\ Rev.\ Lett.\  {\bf 51} (1983) 1135.



\bibitem{ellis-jaffe}
  J.~R.~Ellis and R.~L.~Jaffe,
  Phys.\ Rev.\  D {\bf 9} (1974) 1444
  [Erratum-ibid.\  D {\bf 10} (1974) 1669].


\bibitem{emc}
  J.~Ashman {\it et al.}  [European Muon Collaboration],
  Phys.\ Lett.\  B {\bf 206} (1988) 364.\\
  J.~Ashman {\it et al.}  [European Muon Collaboration],
  Nucl.\ Phys.\  B {\bf 328} (1989) 1.



\bibitem{Abe:1997cx}
  K.~Abe {\it et al.}  [E154 Collaboration],
  Phys.\ Rev.\ Lett.\  {\bf 79} (1997) 26
  [arXiv:hep-ex/9705012].


\bibitem{Abe:1998wq}
  K.~Abe {\it et al.}  [E143 collaboration],
  Phys.\ Rev.\  D {\bf 58}, (1998) 112003
  [arXiv:hep-ph/9802357].


\bibitem{Anthony:1996mw}
  P.~L.~Anthony {\it et al.}  [E142 Collaboration],
  Phys.\ Rev.\  D {\bf 54}, (1996) 6620 
  [arXiv:hep-ex/9610007].


\bibitem{Anthony:1999py}
  P.~L.~Anthony {\it et al.}  [E155 Collaboration],
  Phys.\ Lett.\  B {\bf 458} (1999) 529
  [arXiv:hep-ex/9901006].


\bibitem{Adams:1997tq}
  D.~Adams {\it et al.}  [Spin Muon Collaboration (SMC)],
  Phys.\ Rev.\  D {\bf 56} (1997) 5330
  [arXiv:hep-ex/9702005].


\bibitem{Airapetian:1998wi}
  A.~Airapetian {\it et al.}  [HERMES Collaboration],
  Phys.\ Lett.\  B {\bf 442} (1998) 484
  [arXiv:hep-ex/9807015].

\bibitem{comp.del_sigma}
  V.~Y.~Alexakhin {\it et al.}  [COMPASS Collaboration],
  Phys.\ Lett.\  B {\bf 647} (2007) 8.


\bibitem{compassrho} 
M.~Alekseev {\it et~al}. [COMPASS collaboartion],  Eur.\ Phys.\ J.\ C {\bf 52} (2007) 255.

\bibitem{Abragam:89a}
A.~Abragam, \textit{The Principles of nuclear magnetism} (The Clarendon Press
, Oxford, 1961).

\bibitem{compass}
P. ~Abbon {\it et al.}, Nuclear Instruments and Methods in Physics Research
\textbf{A577} (2007) 455.

\bibitem{geant}
R.~Brun {\it et~al}., CERN Program Library W5013 (1994).

\bibitem{Ingelman:1996mq}
  G.~Ingelman, A.~Edin and J.~Rathsman,
  Comput.\ Phys.\ Commun.\  {\bf 101} (1997) 108
  [arXiv:hep-ph/9605286].

\bibitem{Martin:2006qz}
  A.~D.~Martin, W.~J.~Stirling and R.~S.~Thorne,
  Phys.\ Lett.\  B {\bf 636} (2006) 259
  [arXiv:hep-ph/0603143].


\bibitem{Andersson:89}
B.~Andresson, \textit{The Lund model} (Cambridge Univ. Press
, Cambridge, 1989).


\bibitem{Sjostrand:2000wi}
  T.~Sjostrand, P.~Eden, C.~Friberg, L.~Lonnblad, G.~Miu, S.~Mrenna and E.~Norrbin,
  Comput.\ Phys.\ Commun.\  {\bf 135} (2001) 238
  [arXiv:hep-ph/0010017].

\bibitem{Sulej:2007zz}
  R.~Sulej, K.~Zaremba, K.~Kurek and E.~Rondio,
  Measur.\ Sci.\ Tech.\  {\bf 18} (2007) 2486.

\bibitem{jorg}
J.~Pretz, ``A New Method for Asymmetry Extraction'', COMPASS internal note (2004).

\bibitem{Ageev:2005pq}
  E.~S.~Ageev {\it et al.}  [COMPASS Collaboration],
  Phys.\ Lett.\  B {\bf 633} (2006) 25
  [arXiv:hep-ex/0511028].

\bibitem{Adeva:1999pa}
B.~Adeva {\it et~al}. [SMC], Phys.\ Rev.\ D {\bf 60} (1999) 072004.

\bibitem{Adeva:1998vv}
B.~Adeva {\it et~al}. [SMC], Phys.\ Rev.\ D {\bf 58} (1998) 112001.

\bibitem{Bourrely:1987gp}
C.~Bourrely, J.~Soffer, F.~M. Renard, and P.~Taxil, Phys. Rept. 177 (1989) 319.

\bibitem{Gluck:2000dy}
M.~Gl{\"u}ck, E.~Reya, M.~Stratmann, and W.~Vogelsang, Phys.\ Rev.\ D {\bf 63} (2001) 094005.

\bibitem{Gluck:1998xa}
M.~Gl{\"u}ck, E.~Reya, and A.~Vogt, Eur.\ Phys.\ J.\ C {\bf 5} (1998) 461.

\bibitem{Gluck:2001rn}
M.~Gl{\"u}ck, E.~Reya, and C.~Sieg, Eur.\ Phys.\ J.\ C {\bf 20} (2001) 271.



\bibitem{Adeva:2004dh}
B.~Adeva {\it et~al}. [SMC], Phys.\ Rev.\ D {\bf 70} (2004) 012002.

\bibitem{Airapetian:1999ib}
A.~Airapetian {\it et~al}. [HERMES collaboration], Phys.\ Rev.\ Lett.\ {\bf 84} (2000) 2584.






\end{thebibliography}
\end{document}